\documentclass[%
 reprint,
 superscriptaddress,
nofootinbib,
 amsmath,amssymb,
 aps,prx,
]{revtex4-2}

\usepackage{graphicx}% Include figure files
\usepackage{dcolumn}% Align table columns on decimal point
\usepackage{bm}% bold math
\usepackage{bbm}% extra math
\usepackage{physics}% extra methods
\usepackage{esint}
\usepackage{hepnames}
\usepackage{hepnicenames}  
\usepackage{dsfont}
\usepackage{siunitx}
\usepackage{booktabs}
\usepackage{pgffor}
\usepackage{CJK}

\usepackage[compat=1.1.0]{tikz-feynman}
\usepackage{tikz}
\usepackage{ragged2e}
\usetikzlibrary{shapes.geometric,decorations.markings,plotmarks,positioning,automata,arrows.meta}
\tikzset{every loop/.style={}}
\tikzfeynmanset{
every particle/.style={blue},
every blob/.style={draw=green!40!black, pattern color=green!40!black},
}

\usepackage{hepnames-fix}
\usepackage{comment}
\usepackage{amsmath}
\allowdisplaybreaks[4]

\begin{document}

\preprint{APS/123-QED}

\title{Quantum Steering Geometry at High Energy Particle Colliders}

\author{Juan J. Mejia Alvarez}
\thanks{These authors contributed equally to this work.}
\affiliation{Department of Physics \& Astronomy, Purdue University,
West Lafayette, Indiana 47907, USA}

\author{Andrew J. Wildridge\textsuperscript{*,}}
\email{Contact author: andrew.james.wildridge@cern.ch}
\altaffiliation{Present address: Fermi National Accelerator Laboratory, Batavia, IL 60510, USA}
\affiliation{Department of Physics \& Astronomy, Purdue University,
West Lafayette, Indiana 47907, USA}

\author{Angelo Arisi}
\affiliation{Department of Physics \& Astronomy, Purdue University,
West Lafayette, Indiana 47907, USA}

\author{Juan M. Duarte-Quiros}
\affiliation{Department of Physics \& Astronomy, Purdue University,
West Lafayette, Indiana 47907, USA}

\author{Santosh Bhandari}
\affiliation{Department of Physics \& Astronomy, Purdue University,
West Lafayette, Indiana 47907, USA}

\author{Jingyan Li}
\affiliation{Department of Physics \& Astronomy, Purdue University,
West Lafayette, Indiana 47907, USA}

\author{Giulia Negro}
\affiliation{Department of Physics \& Astronomy, Purdue University,
West Lafayette, Indiana 47907, USA}

\author{Andreas W. Jung}
\affiliation{Department of Physics \& Astronomy, Purdue University,
West Lafayette, Indiana 47907, USA}

\date{\today}% It is always \today, today,

\newlength{\gaincellwidth}
\setlength{\gaincellwidth}{5.0em}

\newcommand{\gain}[2]{%
  \makebox[\gaincellwidth][l]{\ensuremath{#1\, (#2)}}%
}

\newcommand{\bgain}[2]{%
  \makebox[\gaincellwidth][l]{\ensuremath{\mathbf{#1}\,(\mathbf{#2})}}%
}

\newcommand{\cM}{\mathcal{M}}
\newcommand{\cO}{\mathcal{O}}
\newcommand{\cEA}{\ensuremath{\mathcal{E}_{\text{A}|\text{B}}}}
\newcommand{\cEB}{\ensuremath{\mathcal{E}_{\text{B}|\text{A}}}}
\newcommand{\cA}{\ensuremath{\bm{c}_{\text{A}|\text{B}}}}
\newcommand{\cB}{\ensuremath{\bm{c}_{\text{B}|\text{A}}}}
\newcommand{\QA}{\ensuremath{Q_{\text{A}|\text{B}}}}
\newcommand{\QB}{\ensuremath{Q_{\text{B}|\text{A}}}}
\newcommand{\TA}{\ensuremath{\widetilde{T}_{\text{A}|\text{B}}}}
\newcommand{\TB}{\ensuremath{\widetilde{T}_{\text{B}|\text{A}}}}
\newcommand{\gammaA}{\ensuremath{\gamma_{\text{A}}}}
\newcommand{\gammaB}{\ensuremath{\gamma_{\text{B}}}}

\newcommand{\Id}{\mathbb{I}}
\newcommand{\sig}{\ensuremath{\bm{\sigma}}}
\newcommand{\sigA}{\ensuremath{\bm{\sigma}_{\text A}}}
\newcommand{\sigB}{\ensuremath{\bm{\sigma}_{\text B}}}
\newcommand{\BA}{\ensuremath{\bm{B}_{\text A}}}
\newcommand{\BB}{\ensuremath{\bm{B}_{\text B}}}
\newcommand{\OmegaA}{\ensuremath{\hat{\vb{\Omega}}_{\text A}}}
\newcommand{\OmegaB}{\ensuremath{\hat{\vb{\Omega}}_{\text B}}}
\newcommand{\OmegaAbar}{\ensuremath{\overline{\vb{\Omega}}_{\text A}}}
\newcommand{\OmegaBbar}{\ensuremath{\overline{\vb{\Omega}}_{\text B}}}

\newcommand{\Ckk}{\ensuremath{C_{kk}}}
\newcommand{\Crr}{\ensuremath{C_{rr}}}
\newcommand{\Cnn}{\ensuremath{C_{nn}}}
\newcommand{\Crk}{\ensuremath{C_{rk}}}
\newcommand{\Ckr}{\ensuremath{C_{kr}}}

\newcommand{\ctG}{\ensuremath{c_{\Pqt G}}}
\newcommand{\OtG}{\ensuremath{\mathcal{O}_{\Pqt G}}}
\newcommand{\cG}{\ensuremath{c_{G}}}
\newcommand{\OG}{\ensuremath{\mathcal{O}_{G}}}
\newcommand{\cphiG}{\ensuremath{c_{\phi G}}}
\newcommand{\OphiG}{\ensuremath{\mathcal{O}_{\phi G}}}

\newcommand{\QbarL}{\ensuremath{\bar Q_L}}
\newcommand{\QL}{\ensuremath{Q_L}}
\newcommand{\qbarL}{\ensuremath{\Paq_L}}
\newcommand{\qL}{\ensuremath{\Pq_L}}
\newcommand{\tbarR}{\ensuremath{\Paqt_R}}
\newcommand{\tR}{\ensuremath{\Pqt_R}}
\newcommand{\ubarR}{\ensuremath{\Paqu_R}}
\newcommand{\uR}{\ensuremath{\Pqu_R}}
\newcommand{\dbarR}{\ensuremath{\Paqd_R}}
\newcommand{\dR}{\ensuremath{\Pqd_R}}

\newcommand{\nhat}{\ensuremath{\hat{\vb n}}}
\newcommand{\rhat}{\ensuremath{\hat{\vb r}}}
\newcommand{\khat}{\ensuremath{\hat{\vb k}}}

\newcommand{\mt}{\ensuremath{M_{\mathrm{T}}}}
\newcommand{\met}{\ensuremath{\Et^{\mathrm{miss}}}}
\newcommand{\Et}{\ensuremath{E_\mathrm{T}}}
\newcommand{\Lep}{\ensuremath{\ell}}
\newcommand{\Lepp}{\ensuremath{\ell^{+}}}
\newcommand{\Lepm}{\ensuremath{\ell^{-}}}
\newcommand{\E}{\ensuremath{\mathrm{e}}}
\newcommand{\M}{\ensuremath{\mu}}
\newcommand{\ttll}{\ensuremath{\mathrm{t}\bar{\mathrm{t}}\to\ell\ell}}
\newcommand{\ttbar}{\ensuremath{\Pqt \Paqt}\xspace}
\newcommand{\ttlj}{\ensuremath{\mathrm{t}\bar{\mathrm{t}}\to\ell+\rm{jets}}}
\newcommand{\ttdl}{\ensuremath{\mathrm{t}\bar{\mathrm{t}}\to\ell\ell}}
\newcommand{\ttsl}{\ensuremath{\mathrm{t}\bar{\mathrm{t}}\to\ell+\rm{jets}}}
\newcommand{\dy}{\ensuremath{Z/\gamma^*}}
\newcommand{\wjets}{\ensuremath{W+}jets} 
\newcommand{\zjets}{\ensuremath{Z+}jets} 
\newcommand{\tw}{\ensuremath{tW}} 
\newcommand{\vv}{\ensuremath{\mathrm{WW/WZ/ZZ}}}
\newcommand{\vvv}{\ensuremath{\mathrm{WWW/WWZ/WZZ/ZZZ}}}
\newcommand{\mll}{\ensuremath{M_{\Lep\Lep}}}
\newcommand{\mZ}{\ensuremath{M_{\mathrm{Z}}}}
\newcommand{\mtop}{\ensuremath{m_{\Pqt}}}
\newcommand{\mttbar}{\ensuremath{m_{\ttbar}}}
\newcommand{\To}{\ensuremath{\rightarrow}}
\newcommand{\njets}{$N_{\rm{jets}}$}
\newcommand{\lsp}{\ensuremath{\tilde{\chi}_{1}^{0}}}
\newcommand{\TbW}{\ensuremath{\tilde{t}\rightarrow t \tilde{\chi}_{1}^{0}}}
\newcommand{\sieie}{$\sigma_{i\eta i\eta}$}
\newcommand{\ee}{\ensuremath{ee}}
\newcommand{\emu}{\ensuremath{e\mu}}
\newcommand{\mumu}{\ensuremath{\mu\mu}}
\newcommand{\lumi}{\ensuremath{\mathcal{L}}}
\newcommand\Tstrut{\rule{0pt}{2.6ex}}       % Top strut
\newcommand\Bstrut{\rule[-1.2ex]{0pt}{0pt}} % Bottom strut

\newcommand{\thetas}{\ensuremath{\theta^*}}

\newcommand{\PythiaOnly} {{\textsc{Pythia}}} %%%%%%%%%%%%%
\newcommand{\Pythia} {{\textsc{Pythia8}}} %%%%%%%%%%%%%
\newcommand{\Powheg} {{\textsc{Powheg}}} %%%%%%%%%%%%%
\newcommand{\Powhegvtwo}{{\textsc{Powhegv2}}} %%%%%%%%%%%%%
\newcommand{\Herwig} {{\textsc{Herwig}}} %%%%%%%%%%%%%
\newcommand{\MadSpin} {{\textsc{MadSpin}}} %%%%%%%%%%%%%
\newcommand{\Herwigpp} {{\textsc{Herwig++}}} %%%%%%%%%%%%%
\newcommand{\MGaMCatNLO} {MG5\_aMC@NLO(FxFx)} %%%%%%%%%%%%%
\newcommand{\MGaMCatNLOOnly} {MG5\_aMC@NLO} %%%%%%%%%%%%%
\newcommand{\MGMLM} {MG5\_aMC@NLO(MLM)} %%%%%%%%%%%%%

\newcommand{\qqbar}{\ensuremath{\mathrm{q\bar{q}}}\xspace}
\newcommand{\gluglu}{\ensuremath{\Pg\Pg}\xspace}
\newcommand{\pp}{\ensuremath{\Pp\Pp}\xspace}

% Margin notes
\newcounter{mnotecount}[section]
\renewcommand{\themnotecount}{\thesection.\arabic{mnotecount}}
\newcommand{\mnote}[1]%{}
{\protect{\stepcounter{mnotecount}}$^{\mbox{\footnotesize
$%\!\!\!\!\!\!\,
\bullet$\themnotecount}}$ \marginpar{%\color{red}%
\raggedright\tiny\em
$\!\!\!\!\!\!\,\bullet$\themnotecount: #1} }
\newcommand{\diff}[2]{{\color{red}\sout{#1}} {\color{blue}#2}}

\newcommand{\JM}[1]{\mnote{{\bf JM} \textcolor{red}{#1}}}
\newcommand{\JQ}[1]{\mnote{{\bf JQ} \textcolor{blue}{#1}}}
\newcommand{\AAR}[1]{\mnote{{\bf AAR} \textcolor{green}{#1}}}
\newcommand{\AJ}[1]{\mnote{{\bf AJ} \textcolor{magenta}{#1}}}
\newcommand{\AJW}[1]{\mnote{{\bf AJW} \textcolor{magenta}{#1}}}

% Keep an original \bm without automatic vector
\let\oldbm\bm  

% Redefine \bm globally to always add \vec
\renewcommand{\bm}[1]{\vec{\oldbm{#1}}}
\newcommand{\invfb}{fb$^{-1}$}
\newcommand{\invpb}{pb$^{-1}$}

\newcolumntype{X}[1]{D{,}{\,\pm\,}{#1}}

\begin{abstract}

We formulate collider observables based on quantum steering ellipsoids (QSEs) for reconstructed bipartite systems of spin-$1/2$ particles. A collider spin density matrix defines a two-qubit state, while its QSE gives the geometry of conditional states accessible through local measurements. This makes the ellipsoid a direct probe of the quantum properties of fundamental particles, encoding polarization, spin correlation anisotropy, accessible-state volume, and the orientation of the dominant correlation axes. Using top-quark pair production as a benchmark process, we show how QSE observables organize the Standard Model spin state, probe entanglement, steerability, and Bell-nonlocality criteria, and use an expected precision metric to assess sensitivity to non-local correlations in the boosted central region. We show that different dimension-six operators generate distinctive QSE deformations, and provide a geometric interpretation of quantum information observables in high-energy particle physics. Quantum steering geometry therefore provides a unified framework for particle collider tomography, quantum information diagnostics, and precision searches for physics beyond the Standard Model with applications spanning the HL--LHC and future lepton, muon, flavor, and electron-ion collider programs.

\end{abstract}

\maketitle

\section{Introduction}

Spin correlations and quantum information observables in relativistic scattering processes provide a unique window into the quantum structure of fundamental interactions. In recent years, measurements of top quark-antiquark spin correlations~\cite{ATLAS:2019zrq, TOP-11-005_paper, CMS:2019nrx, CMS:2024zkc, ATLAS:2016bac, ATLAS:2012ao, Lemmer:2014sxa,D0:2011kcb}, spin density matrices, quantum tomography~\cite{CMS:2019nrx, CMS:2024zkc}, and entanglement~\cite{ATLAS:2023fsd, CMS:2024pts, CMS:2024zkc} at the Large Hadron Collider (LHC) have demonstrated that collider experiments can directly access quantum information properties of elementary-particle production processes. These developments establish high-energy particle colliders as experimental laboratories for quantum information science (QIS), where reconstructed spin density matrices provide experimentally accessible two-qubit states that encode the underlying dynamics of particle interactions. Although existing studies have primarily focused on entanglement measures~\cite{Afik:2020onf, Fabbrichesi:2021npl, Severi:2021cnj, Aoude:2022imd,Severi:2022qjy, Afik:2022kwm, Aguilar-Saavedra:2022uye, Han:2023fci, Dong:2023xiw, Cheng:2024btk, Aguilar-Saavedra:2022wam, Ashby-Pickering:2022umy, Aguilar-Saavedra:2022mpg,  Fabbrichesi:2023cev,  Barr:2024djo, Afik:2025ejh} and spin correlation observables~\cite{Barr:2021zcp, Barr:2022wyq, Altakach:2022ywa, Cheng:2023qmz, Han:2024ugl, White:2024nuc, Aoude:2025jzc}, the full geometric structure of the reconstructed quantum state remains largely unexplored.

\begin{figure}[!ht]
    \centering
    \includegraphics[width=0.9\linewidth]{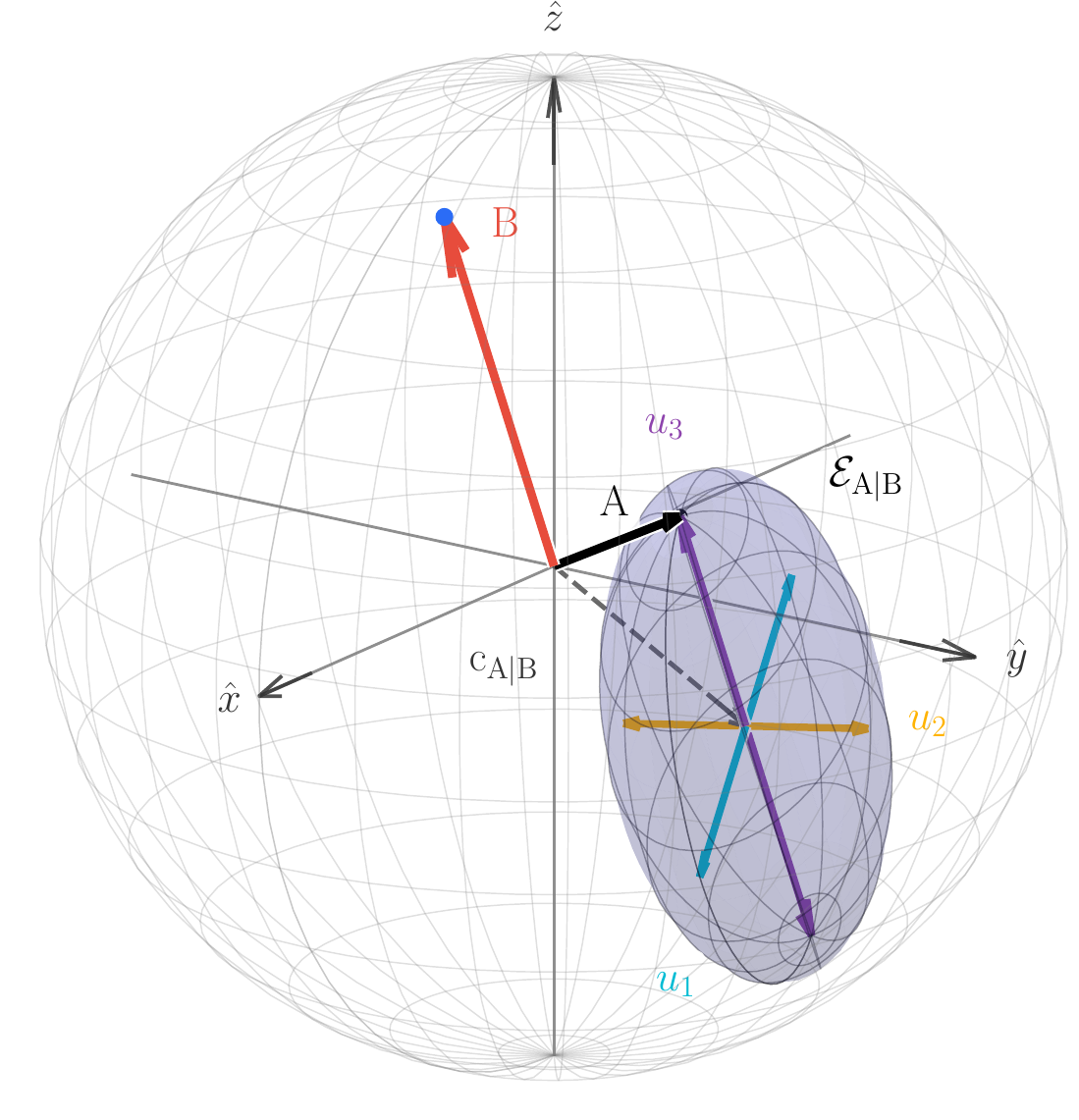}
    \caption{Example of a quantum steering ellipsoid $\cEA$ embedded in the Bloch sphere. The center $\cA$ is shown, as well as the semi-axes $u_1$, $u_2$, and $u_3$. A local positive operator-valued measurement outcome on subsystem $\text B$ with a unit Bloch vector on the surface of the Bloch sphere yields a state on subsystem $\text A$ that lies on the surface of the quantum steering ellipsoid.}
    \label{fig:EllipsoidInBlochSphere}
\end{figure}

In this work, we formulate a new class of collider observables based on quantum steering ellipsoids (QSEs)~\cite{Jevtic:2014icx, Jevtic:2015epl} for reconstructed bipartite systems of spin-$1/2$ particles, see Fig.~\ref{fig:EllipsoidInBlochSphere}. For any pair of spin-$1/2$ particles, identified as Alice's (A) and Bob's (B) subsystems, the spin production matrix $R$ can be expanded in the Pauli basis, commonly referred to as ``Fano representation or decomposition"~\cite{Fano, Bernreuther2015, Han:2023fci}
\begin{equation}\label{eq:Spin density matrix}
    R:=\tilde A\mathds{1}_2\otimes\mathds{1}_2
    +\tilde{B}^i_{\text A}\sigma_i\otimes\mathds{1}_2
    +\tilde{B}^i_{\text B}\mathds{1}_2\otimes\sigma_i
    +\tilde C_{ij}\sigma^i\otimes\sigma^j,
\end{equation}
\noindent where $\{\sigma_i\}_{i=1,2,3}$ are the Pauli matrices~\cite{Pauli1927}. The corresponding normalized two-qubit state is
\begin{equation}\label{eq:Quantum State From R}
    \rho_{\text{AB}}=\frac{R}{\Tr(R)}.
\end{equation}
The vectors $\bm{\tilde{B}}_{\text{A}(\text{B})}$ encode the single-particle polarizations, while $\tilde C$ encodes the spin correlations~\cite{Bernreuther2015}. After normalizing by $\tilde A$:
\begin{equation}
    \label{eq:fano_coeffs_norm}
    \bm B_{\text{A}(\text{B})}:=\frac{\bm{\tilde{B}}_{\text{A}(\text{B})}}{\tilde A}\,,
    \qquad
    C:=\frac{\tilde C}{\tilde A}\,,
\end{equation}
the polarization vectors $(\BA,\BB)$ together with the correlation matrix $C$ fully determine the two-qubit spin state. 

The spin production matrix can be computed from the polarized scattering amplitudes for a process $a_0+\cdots+a_n\to A+B$~\cite{Bernreuther2015, Afik:2020onf, Afik:2022dgh, Aoude:2022imd, Han:2023fci},
\begin{equation}
\label{eq: R matrix from scattering}
R_{a\bar a,b\bar b}=
\overline{\sum_{\rm initial}} \mathcal M_{a\bar a} \mathcal M^*_{b \bar b},
\end{equation}
\noindent where ${\mathcal M}$
is the scattering matrix element for producing $A$ and $B$ with spin indices $ab$ and $\bar{a}\bar{b}$, respectively, and the sum denotes the average over initial-state spin and color degrees of freedom. In this sense, collider spin tomography turns perturbative scattering amplitudes into experimentally accessible information about the spin quantum state of the produced two-particle subsystem.

In collider applications, the QSE recasts the Fano coefficients provided in Eq.~\eqref{eq:Spin density matrix} into observables associated with the center, semi-axes, orientation, surface area, and volume of an ellipsoid. These observables are not independent of the reconstructed density matrix; rather, they provide nonlinear and geometrically interpretable combinations of the polarization vectors and spin correlation matrix. Thus, the QSE provides a general geometric descriptor for arbitrary two-qubit density matrices, independent of the underlying particle flavor or production mechanism. This makes them useful both as probes of the quantum structure of collider-produced bipartite states and as complementary observables in searches for physics beyond the Standard Model (BSM). 

We demonstrate this program in the top quark-antiquark system. Section~\ref{sec:collider_steering} introduces the QSE formalism, derives its relation to differential cross sections in collider physics, and shows how coarse graining enters the reconstruction in realistic collider applications. This section also defines the QSE observables used in the subsequent phenomenological studies.
 
Section~\ref{sec:ttbar_qse_reconstruction} describes the computational setup and validates the QSE reconstruction using next-to-leading order (NLO) quantum chromodynamics (QCD) simulated top quark-antiquark ($\ttbar$) events with spin-analyzing decays. We compare the resulting ellipsoids with an independent leading-order (LO) analytic calculation, validating the setup used to study deformations in the Standard Model Effective Field Theory (SMEFT).
In Section~\ref{sec:qse_bell_ineq}, we observe that QSE-based Bell observables provide sensitivity to Bell-nonlocal spin correlations similar to that of Fano-based Bell observables. We perform a geometric analysis of the QSEs in the SMEFT in Section~\ref{sec:smeft_qse_deformations}, allowing a direct comparison between traditional tomographic QIS observables and QSE observables. We find significant improvements in expected limits on new-physics contributions when including QSE-based observables.

\section{Collider steering ellipsoids}
\label{sec:collider_steering}

Steering, first discussed by Schr\"{o}dinger~\cite{Schrodinger:2008pyl} in response to the Einstein--Podolsky--Rosen (EPR) Gedankenexperiment~\cite{EPR}, captures the fact that measurements performed on one subsystem can condition the state assigned to the other. For a bipartite system of spin-$1/2$ particles, varying the local measurement on subsystem $\text B$ generates a family of conditional states for subsystem $\text A$. In the Bloch representation, the endpoints of the corresponding conditional Bloch vectors form an ellipsoid inside the Bloch sphere: the quantum steering ellipsoid~\cite{Jevtic:2014icx,Jevtic:2015epl}. 

This construction extends beyond visualization. The geometric properties of the QSE have been shown to encode Bell nonlocality~\cite{Milne:2014mtq,Quan:2016nis}, EPR steerability~\cite{Jevtic:2015epl}, entanglement~\cite{Jevtic:2014icx}, and quantum discord~\cite{Nassajpour:2018jeu,Jevtic:2014icx}. The QSE has also proven useful in characterizing quantum coherence~\cite{Hu:2016ksu}, quantum phase transitions~\cite{Rosario:2023dwu}, and monogamy of entanglement~\cite{Cheng:2016klf,Zhang:2019hsk,Milne:2014nsw}, among other phenomena~\cite{Xu:2023fiz}. These studies establish the QSE as a versatile geometric framework for quantum correlations within quantum information science and atomic, molecular, and optical (AMO) physics. Motivated by these developments, we investigate whether the same geometry can be translated into experimentally accessible observables for reconstructed quantum states at particle colliders. 

This construction is especially natural for collider spin tomography. The experimentally reconstructed spin state is generally mixed, either because unobserved degrees of freedom have been averaged over or because the state has been integrated over a finite region of phase space. For mixed two-qubit states, the set of conditional states is not described by a single spin correlation coefficient or by a single entanglement witness. Instead, the Fano coefficients determine an entire ellipsoid of conditional spin states. The QSE is therefore not independent of the density matrix; rather, it provides a nonlinear geometric reorganization of the same polarization vectors and spin correlation matrix into experimentally accessible observables. While mathematically equivalent to the underlying spin-density matrix, this representation naturally gives rise to a new class of geometric collider observables whose independent phenomenological information content is quantified in the following sections. We first review the QSE construction, show how its points are accessed through collider differential cross sections, and define the QSE and QIS observables used throughout this work.

\subsection{Steering as local-outcome conditioning}\label{subsec: Steering}

To describe the most general local quantum measurements, we employ positive operator-valued measures (POVMs). Each measurement outcome is represented by a positive operator (a POVM element, or effect), which in the Pauli basis takes the form
\begin{equation}
\label{eq:qubit_povm}
E(\bm m)
=
m_0
\left(
\mathds{1}_2+\bm m\cdot\bm\sigma
\right).
\end{equation}
Here $m_0\geq0$ is the normalization and $\bm m$ is the Bloch vector associated with the outcome, with $\|\bm m\|\leq1$. Rank-one projective effects have $\|\bm m\|=1$, while non-rank-one or noisy outcomes have $\|\bm m\|<1$. A POVM is then a collection of such effects, $\{E(\bm{m}_i)\}$, whose elements are positive Hermitian operators that satisfy the completeness relation, $\sum_i E(\bm{m}_i)=\mathbb{I}$. In what follows, we take Bob's measurement outcome as the conditioning outcome, so that the effect $E(\bm m)$ acts as $\mathds{1}_2\otimes E(\bm m)$ and defines a conditional state for Alice.

To evaluate this conditional state, we write the normalized spin density matrix from Eq.~\eqref{eq:Quantum State From R} in the Fano--Bloch form~\cite{Fano},
\begin{equation}\label{eq:Fano Decomposition two qubit state}
    \rho_{\text{AB}}=\frac{1}{4}\left[\mathds{1}_4+\BA\cdot\sigA+ \BB\cdot\sigB+ \sigA\cdot C\cdot\sigB
    \right],
\end{equation}
\noindent $\sigA=\sig\otimes\mathds{1}_2$, $\sigB=\mathds{1}_2\otimes\sig$. The reduced density matrices are

\begin{align}
    &\rho_{\text A}=\Tr_{\text B}[\rho_{\text{AB}}]=\frac{1}{2}\left(\mathds{1}_2+\BA\cdot\sig\right),\\
    &\rho_{\text B}=\Tr_{\text A}[\rho_{\text{AB}}]=\frac{1}{2}\left(\mathds{1}_2+\BB\cdot\sig\right).
    \label{eq:reduced_density_matrices_main}
\end{align}
Thus, $\BA$ and $\BB$ are the Bloch vectors of the individual subsystems. After conditioning on the effect $E(\bm m)$ on Bob, using Eq.~\eqref{eq:qubit_povm} Alice's state becomes~\cite{Jevtic:2014icx}
\begin{align}\label{eq:partialTraceAliceSteering}
    \rho_{\text A}^{E_m} &=\frac{\Tr_{\rm B}\left[(\mathds 1_2\otimes E_m)\rho_{\rm AB}\right]}{\Tr\left[(\mathds 1_2\otimes E_m)\rho_{\rm AB}\right]}\nonumber\\&= \frac{1}{2}\left[\mathds{1}_2 + \frac{\left(\BA + C \bm{m}\right) }{1 + \BB \cdot \bm{m}}\cdot \sig \right].
\end{align}
The expression multiplying $\sig$ is the Bloch vector of Alice's conditional state. Varying Bob's measurement outcome over all allowed $\bm m$ therefore gives the set
\begin{equation}
    \label{eq:qse}
    \cEA = 
    \left\{ 
        \frac{\left(\BA + C \bm{m}\right)}{1 + \BB \cdot \bm{m}} 
        : \|\bm{m} \| \leq 1
    \right\}.
\end{equation}
The resulting set, known as the quantum steering ellipsoid (QSE), forms an ellipsoid contained within the Bloch sphere~\cite{FrankVerstraetePhdThesis,Jevtic:2014icx}. Originally introduced in quantum information as a geometric representation of two-qubit states~\cite{Jevtic:2014icx}, the QSE has primarily been studied in quantum information~\cite{Milne:2014nsw, Cheng:2016klf} and quantum-optical settings~\cite{Zhang:2019hsk, Xu:2023fiz}. Here, we formulate its geometric characteristics as observables of reconstructed collider spin states and show how to access them through the angular distributions of spin-analyzing decay products. The QSE can be written in the standard quadratic form
\begin{equation}
    \label{eq:QSEdef}
    \left(\bm{x} - \cA\right)^{\top}
    \QA^{-1}
    \left(\bm{x} - \cA\right) \leq 1\,,
\end{equation}
where $\cA$ is the center of the ellipsoid, $\QA$ is the real symmetric matrix that determines its shape and orientation, and $\top$ denotes transposition. These quantities are fixed by the Fano coefficients as given in Ref.~\cite{Jevtic:2014icx}
\begin{align}
    \label{eq:fanoQSECent}
    \cA &= \gammaB^2
    \left(\BA-C\,\BB\right)\,,\\
    \label{eq:transformationDef}
    \TA &= \gammaB
    \left(C - \BA \BB^\top \right)
    \left(\mathds{1}_3 + \frac{\gammaB - 1}{\BB^2}\BB\BB^\top \right), \\
    \QA &= \TA\TA^{\top}~,
\end{align}
where $\gammaB=\left(1-\|\BB\|^2\right)^{-1/2}$ comes from the standard QSE construction~\cite{Avron_2007SLOCC}. 

The subscript $\text{A} | \text{B}$ refers to Alice's QSE obtained by conditioning on Bob's local outcomes. Bob's QSE can be obtained by exchanging $\text{A}\leftrightarrow\text{B}$ and $C \leftrightarrow C^{\top}$. Unless otherwise stated, we focus on Alice's QSE and suppress the subscript $\text A|\text B$ in the following sections.

\subsection{Collider realization}

Many collider processes, including top-quark pairs, bottom-quark pairs, Higgs and Z-boson decays to $\tau^+\tau^-$, produce unstable spin-$1/2$ particle pairs whose decay products retain information about the underlying quantum state~\cite{Fabbrichesi:2022ovb,Afik:2020onf,Afik:2025grr,CMS:2024uqa}. This provides a natural experimental realization of the QSE formalism, with local quantum measurements encoded in the angular distributions of spin-analyzing decay products.

For unstable particles produced at particle colliders, the POVM effect $E(\bm m)$ is realized through the angular distribution of a spin-analyzing decay product. 
Under the narrow-width approximation~\cite{Bernreuther2015}, the production and decay stages factorize without tracing over the intermediate spin degrees of freedom. Consequently, the angular distributions of spin-analyzing decay products directly encode the polarization vectors and spin correlations generated in the hard scattering, providing the experimental realization of the local quantum measurements required for the QSE construction.

For a selected decay product emitted along the direction $\hat{\vb{\Omega}}_{\text X}$ in the rest frame of the parent particle $X=\text A, \text B$, the angular measurement is described by the POVM density~\cite{Ashby-Pickering:2022umy}
\begin{equation}\label{eq:spin analyzer POVM}
    \frac{\dd E_{\text X}}{\dd^2 \Omega_{\text X}}(\hat{\vb{\Omega}}_{\text X}) = \frac{1}{4\pi}
    \left(
        \mathds{1}_2+\kappa_{\text X}\hat{\vb{\Omega}}_{\text X}\cdot\bm\sigma
    \right).
\end{equation}
The selected decay product acts as the spin analyzer, with $\kappa_{\text X}$ determining the strength of its correlation with the parent spin.

In the notation of Eq.~\eqref{eq:qubit_povm}, an infinitesimal analyzer outcome corresponds to the Bloch vector $\bm m=\kappa_{\text X}\hat{\vb{\Omega}}_{\text X}$. Thus, an ideal analyzer with $|\kappa|=1$ gives a unit-norm outcome, while $|\kappa|<1$ gives an effective outcome measurement inside the Bloch sphere.

Normalization in Eq.~\eqref{eq:spin analyzer POVM} ensures completeness, 
\begin{equation} 
    \int_{S^2} 
    \dd^2\Omega\,
    \frac{\dd E}{\dd^2\Omega}(\hat{\vb{\Omega}}) = 
    \mathds{1}_2, 
\end{equation} 
where $S^2$ is the unit sphere. Combining the angular distributions from both decay chains then reconstructs the production density matrix $\rho_{\text{AB}}$, providing direct access to the polarization vectors and spin correlations.

Applying the Born rule to the analyzer POVMs together with the Fano representation of Eq.~\eqref{eq:Fano Decomposition two qubit state} yields the normalized four-fold differential cross section distribution in terms of the angular directions of the subsystems,

\begin{align}
\label{eq:Four fold differential cross section}
\frac{1}{\sigma} 
\frac{\dd^4\sigma}{\dd^2\Omega_{\text A}\dd^2\Omega_{\text B}}
= 
\frac{1}{(4\pi)^2}
\bigg[
1
&+
\kappa_{\text A}\BA\cdot\OmegaA
+
\kappa_{\text B}\BB\cdot\OmegaB
\nonumber\\
&+
\kappa_{\text A}\kappa_{\text B}
\OmegaA^{\top}C\OmegaB
\bigg].
\end{align}

Equation~\eqref{eq:Four fold differential cross section} establishes the experimental connection between collider measurements and the Fano representation: fitting the measured angular distribution determines the polarization vectors $\BA$ and $\BB$, and the spin correlation matrix $C$, which form the input to the QSE construction. 

The connection between collider spin tomography and the QSE is obtained by conditioning on Bob's analyzer direction. Using Eq.~\eqref{eq:Four fold differential cross section}, we first get the marginal distribution for Bob's analyzer as
\begin{equation}
\label{eq:twofold_angular_xsec}
    \frac{1}{\sigma}
    \frac{\dd^2\sigma}{\dd^2\Omega_{\text B}}
    =
    \frac{1}{4\pi}
    \left(
        1+\kappa_{\text B}\BB\cdot\OmegaB
    \right).
\end{equation}
Using Bayes' theorem to condition on Bob's analyzer direction yields the conditional angular distribution
\begin{align}
\label{eq:conditional_angular_distribution}
    \left.
    \frac{1}{\sigma_{\Omega_{\text B}}}
    \frac{\dd^2\sigma_{\Omega_{\text B}}}{\dd^2\Omega_{\text A}}
    \right|_{\OmegaB}
    =
    \frac{1}{4\pi}
    \left[
        1
        +
        \kappa_{\text A}
        \frac{
            \BA+\kappa_{\text B}C\OmegaB
        }{
            1+\kappa_{\text B}\BB\cdot\OmegaB
        }
        \cdot\OmegaA
    \right].
\end{align}
This has the form of a single spin-$1/2$ angular distribution with effective polarization
\begin{equation}
\label{eq:steering_map_collider}
    \BA(\bm m')
    =
    \frac{\BA+C\bm m'}{1+\BB\cdot\bm m'},
    \qquad
    \bm m'
    =
    \kappa_{\text B}\OmegaB .
\end{equation}
Each analyzer direction therefore selects a point on Alice's steering ellipsoid. The spin-analyzing power rescales the accessible Bloch vectors from $\OmegaB$ to $\kappa_{\text B}\OmegaB$: for an ideal analyzer ($|\kappa_{\text B}|=1$), varying $\OmegaB$ traces the QSE surface, whereas for $|\kappa_{\text B}|<1$ only the ellipsoid interior is accessible.

Now, Equation~\eqref{eq:steering_map_collider} is precisely the QSE steering map~\cite{Jevtic:2015epl, Jevtic:2014icx}, here realized through collider spin analyzers. The corresponding collider realization of this construction has previously been explored in the context of quantum steering and discord~\cite{Afik:2022dgh}. In the present work, however, we use this steering map as the starting point for constructing and studying a new family of geometric collider observables based on the full quantum steering ellipsoid.

\subsection{Experimental realization of the steering POVM}
\label{subsec:coarse-grained-steering}

For any practical collider realization, the steering map must be formulated in terms of experimentally reconstructed observables rather than continuous analyzer directions. The steering map provided in Eq.~\eqref{eq:steering_map_collider} assumes that Bob's analyzer direction is resolved exactly. In practice, the continuous direction $\OmegaB$ is reconstructed with finite angular resolution and grouped into solid-angle bins. This replaces the continuous analyzer POVM by a coarse-grained POVM whose outcomes are the bin labels.

Let $\Delta\Omega_{\text B}^{(j)}\subset S^2$ denote the $j^{\text{th}}$ bin in Bob's analyzer direction, with solid angle $\Delta_j$.

The corresponding POVM element is obtained by integrating the POVM density over the bin $j$,
\begin{align}
\label{eq:binned_povm_element}
    E_{\text B}^{(j)}
    &\equiv
    \int_{\Delta\Omega_{\text B}^{(j)}}
    \dd^2\Omega_{\text B}\,
    \frac{\dd E_{\text B}(\OmegaB)}{\dd^2\Omega_{\text B}}
    \nonumber\\
    &=
    \frac{\Delta_j}{4\pi}
    \left(
        \mathds{1}_2
        +
        \bm m'_j\cdot\bm\sigma
    \right),
    \quad
    \bm m'_j
    \equiv
    \kappa_{\text B}\overline{\Omega}_{\text B}^{(j)}.
\end{align}
Here 
\begin{equation}
\label{eq:binned_effective_bloch_vector}
    \overline{\Omega}_{\text B}^{(j)}
    \equiv
    \frac{1}{\Delta_j}
    \int_{\Delta\Omega_{\text B}^{(j)}}
    \dd^2\Omega_{\text B}\,
    \OmegaB
\end{equation}
is the average analyzer direction inside the bin. Although each event has $\|\OmegaB\|=1$, the average direction satisfies $\|\overline{\Omega}_{\text B}^{(j)}\|<1$ for a finite bin. Finite angular binning, therefore, acts in the same way as a non-ideal spin analyzer: it reduces the length of the effective POVM Bloch vector. The effect of a bin is therefore captured by evaluating the steering map at $\bm m'_j$ rather than at $\kappa_{\text B}\OmegaB$.

For experimental applications, we formulate the conditional angular distribution in terms of finite solid-angle bins. The resulting distribution retains the same functional form,
\begin{align}
\label{eq:binned_conditional_distribution}
    \frac{1}{\sigma_j}
    \frac{\dd^2\sigma_j}{\dd^2\Omega_{\text A}}
    &=
    \frac{1}{4\pi}
    \left[
        1+\kappa_{\text A}\BA^{(j)}\cdot\OmegaA
    \right],
    \\
\label{eq:binned_steering_map}
    \BA^{(j)}
    &\equiv
    \BA(\bm m'_j)
    =
    \frac{\BA+C\bm m'_j}{1+\BB\cdot\bm m'_j}.
\end{align}
Remarkably, coarse-graining preserves the functional form of the collider steering map. Consequently, the QSE can be reconstructed directly from experimentally binned angular distributions by replacing the continuous analyzer direction with the corresponding effective bin POVM Bloch vector $\bm m'_j$.

In the combined limit $\kappa_{\text B}\to1$ and $\Delta\Omega_{\text B}^{(j)}\to0$, one has $\overline{\Omega}_{\text B}^{(j)}\to\OmegaB^{(j)}$, and the reconstructed points approach the nominal QSE surface. 

This inward displacement also gives a simple estimate of the volume suppression caused by angular binning. For a partition of the unit sphere into $N=N_\vartheta N_\varphi$ bins, we define 
\begin{equation}
\label{eq:coarse_grained_volume_factor}
    V'(N_\vartheta,N_\varphi)
    \equiv
    \left[
        \frac{1}{N}
        \sum_{j=1}^{N}
        \left\|
            \overline{\Omega}_{\text B}^{(j)}
        \right\|
    \right]^3.
\end{equation}
For an ellipsoid obtained as an affine image of the Bloch sphere, the corresponding coarse-grained volume is approximated by
\begin{equation}
\label{eq:coarse_grained_ellipsoid_volume}
    V_{\rm ell}^{\rm coarse}
    \simeq
    V'(N_\vartheta,N_\varphi)\,
    V_{\rm ell}.
\end{equation}
The factor $V'(N_\vartheta,\,N_\varphi)$ isolates the geometric effect of finite angular binning. It is independent of the ellipsoid parameters and approaches unity as the angular binning is refined.
\subsection{Quantum information and QSE observables}
\label{subsec:qis_qse_observables}

The reconstructed quantum state obtained through collider spin tomography can be characterized by two complementary families of observables. Conventional quantum information observables quantify properties such as entanglement, steering, and Bell nonlocality. In contrast, we show that the geometry of the quantum steering ellipsoid naturally defines a complementary family of collider observables through its center $\bm c$ and shape matrix $Q$. Together, these observables form the basis for the phenomenological studies presented in the remainder of this paper.\\

\noindent\textbf{Entanglement witnesses: $\Delta_\pm$.}

For two qubits, the Peres--Horodecki criterion provides a necessary and sufficient condition for separability~\cite{bib:PhysRevLett.77.1413,Horodecki:1996qk}. In the collider spin bases considered here, convenient sufficient entanglement witnesses are constructed from linear combinations of the diagonal spin correlations~\cite{Aguilar-Saavedra:2022uye,Afik:2020onf},
\begin{align}
    \Delta_+:=&-C_{11}+|C_{33}+C_{22}|-1,
    \label{eq:Delta_plus} \\
    \Delta_-:=&C_{11}+|C_{33}-C_{22}|-1.
    \label{eq:Delta_minus}
\end{align}
A positive value of either witness certifies entanglement. The absolute values select among linear signed combinations of the diagonal correlations, which can be measured directly from the opening-angle distributions.

\vspace{0.5em}

\noindent\textbf{Opening-angle observables: $D$ and $D_a$.}

The opening-angle observables provide experimentally accessible linear combinations of the spin correlation matrix. They are constructed by applying the reflection $P_i$ to one analyzer direction,
\begin{align}
\label{eq:opening_angles}
    &\cos\varphi_i
    =
    \OmegaA^{\top}
    P_i
    \OmegaB,
    &P_0=\mathds{1},
    \\
    &(P_i)_{jk}
    =
    \left(1-2\delta_{ij}\right)\delta_{jk},
    \nonumber
\end{align}
with $i\in\{1,2,3\}$. The corresponding observables are
\begin{equation}
\label{eq:D_observables}
    D_a
    =
    \Tr(P_aC)/3,
    \qquad
    D\equiv D_0=\Tr C/3.
\end{equation}
Thus, $D$ is the opening-angle correlation for aligned analyzer bases, while $D_a$ is obtained after reflecting one analyzer axis. The quantum state is entangled when $D<-1/3$ or $D_a>1/3$.

\vspace{0.5em}

\noindent\textbf{EPR steering indicator: $\hat S$.}

EPR steering quantifies whether measurements on one subsystem can remotely prepare conditional states of the other beyond any local-hidden-state description~\cite{Bell,Quan:2016nis}. For states with vanishing local polarizations, steerability is characterized by~\cite{Nguyen:2016uga,Jevtic:2015epl,Afik:2022dgh}
\begin{equation}
\label{eq:SteerCondition}
    \hat S
    \equiv
    \frac{1}{2\pi}
    \oiint_{S^2}
    \dd^2\hat n\,
    \sqrt{\hat n^{\top}C^{\top}C\hat n}.
\end{equation}
The condition $\hat S>1$ certifies EPR steerability.

\vspace{0.5em}

\noindent\textbf{Bell nonlocality: $\mathcal I_{\rm CHSH}$.}

Bell nonlocality probes whether the reconstructed quantum state admits a local-hidden-variable description and is characterized by the CHSH inequality. Given a set of two spin observables per observer,
\[
A_\alpha=\hat a_\alpha\cdot\bm\sigma,
\qquad
B_\beta=\hat b_\beta\cdot\bm\sigma,
\]
with $\alpha,\beta\in\{1,2\}$, the CHSH inequality reads
\begin{equation}
\label{eq:chsh_ineq}
\left|
\langle A_1B_1\rangle
+
\langle A_1B_2\rangle
+
\langle A_2B_1\rangle
-
\langle A_2B_2\rangle
\right|
\leq2.
\end{equation}
For a two-qubit state with spin correlation matrix $C$, the maximal CHSH value is given by the Horodecki criterion~\cite{HORODECKI1995340},
\begin{equation}
    2\sqrt{u_1+u_2}\leq2,
\end{equation}
where $u_1$ and $u_2$ are the two largest eigenvalues of $C^{\top}C$. We define the Horodecki CHSH indicator
\begin{equation}
\label{eq:BellMaxCondition}
    \mathcal I_{\rm CHSH}
    \equiv
    u_1+u_2-1.
\end{equation}
Positive values of $\mathcal I_{\rm CHSH}$ indicate violation of the CHSH inequality.

\vspace{0.5em}

\noindent\textbf{Collider CHSH indicator: $\mathcal I_{ij}^{(\pm)}$.}

The Horodecki criterion is optimal, but its eigenvalues depend nonlinearly on the measured Fano coefficients. Propagating uncertainties through Eq.~\eqref{eq:BellMaxCondition} may therefore introduce biases in collider measurements~\cite{Severi:2021cnj}. We therefore also consider the sufficient CHSH criterion~\cite{Aguilar-Saavedra:2022uye}
\begin{equation}
\label{eq:bell_cii_cjj}
    |C_{ii}\pm C_{jj}|
    \leq
    \sqrt2,
    \qquad
    i\neq j,
\end{equation}
with the corresponding indicator
\begin{equation}
\label{eq:bell_cii_cjj_indic}
    \mathcal I_{ij}^{(\pm)}
    \equiv
    |C_{ii}\pm C_{jj}|-\sqrt2.
\end{equation}
Positive values of $\mathcal I_{ij}^{(\pm)}$ indicate violation of the CHSH inequality. Violation of Eq.~\eqref{eq:bell_cii_cjj} implies violation of Eq.~\eqref{eq:BellMaxCondition}, but not conversely.\\

We now introduce the complementary family of geometric collider observables derived from the center $\bm c$ and the shape matrix $Q$ of the quantum steering ellipsoid. Unlike the preceding quantum information observables, these observables directly characterize the geometry of the reconstructed quantum state.\\

\noindent\textbf{Geometric QSE parameters: $Q$ and $\bm c$}

The QSE observables are defined from the center $\bm c$ and the shape matrix $Q$, whose components in the spin basis are denoted by $c_i$ and $Q_{ij}$. 

\vspace{0.5em}

\noindent\textbf{QSE semi-axes and principal directions: $s_i$ and $\bm v_i$.}

The semi-axes and principal directions are obtained from the eigensystem
\begin{equation}
    Q\bm v_i=q_i\bm v_i,
    \qquad
    q_i=s_i^2,
    \qquad
    q_1\geq q_2\geq q_3,
\end{equation}
so that $s_i$ are the semi-axis lengths and $\bm v_i$ are the principal directions of the ellipsoid.

\vspace{0.5em}

\noindent\textbf{QSE Bell indicators:
$\mathcal I_{\rm QSE}^{(1)}$ and $\mathcal I_{\rm QSE}^{(2)}$.}\\
For states with vanishing local polarizations, the QSE shape matrix satisfies $Q=CC^{\top}$. Therefore, the nonzero eigenvalues of $Q$ coincide with those of $CC^{\top}$.
From the definition of the shape matrix and Eqs.~\eqref{eq:BellMaxCondition} and~\eqref{eq:bell_cii_cjj}, we can rewrite the CHSH condition in terms of the eigenvalues of the matrix $Q$ as
\begin{equation}
    \label{eq:qse_bell_ineq}
    \mathcal I_{\rm QSE}^{(1)}
    \equiv
    s_1^2+s_2^2-1.
\end{equation}
Positive values of \(\mathcal I_{\rm QSE}^{(1)}\) indicate that the reconstructed QSE geometry lies in the CHSH-violating region. Geometrically, this condition requires the hypotenuse formed from the two longest semi-axes to exceed unity.

In a similar spirit to Eq.~\eqref{eq:bell_cii_cjj_indic}, the related sufficient indicator,
\begin{equation}
    \label{eq:suff_qse_bell_ineq}
    \mathcal I_{\rm QSE}^{(2)}
    \equiv
    s_1+s_2-\sqrt{2},
\end{equation}
indicates that the state lies in the CHSH-violating region for positive values. Geometrically, this condition requires that the total length of the two longest semi-axes exceed $\sqrt{2}$. These conditions are exactly equivalent to Eqs.~\eqref{eq:BellMaxCondition} and~\eqref{eq:bell_cii_cjj} in the absence of polarization.

\vspace{0.5em}

\noindent\textbf{QSE entanglement criterion: $\xi$.}

We also use the QSE form of the Peres--Horodecki criterion~\cite{Jevtic:2014icx, Jevtic:2015epl}. Writing the ellipsoid center as $\bm c=c\hat{c}$, define
\begin{align}
    \label{eq:ellipse_ppt}
    \xi =& c^4-2c^2(1-\Tr Q+2\hat{c}^{\top}Q\hat{c})+h(Q)<0,\\
    h(Q):=&1-8\sqrt{\det Q}+2 \Tr(Q^2)-(\Tr Q)^2-2\Tr Q. \nonumber
\end{align}
The condition $\xi<0$ certifies entanglement.

\vspace{0.5em}

\noindent\textbf{QSE volume: $V_{\rm ell}$.}

The same semi-axes also determine the volume of the ellipsoid,
\begin{equation}
\label{eq:qse_volume}
    V_{\rm ell}
    = \frac{4\pi}{3}\sqrt{\det Q}
    = \frac{4\pi}{3}s_1s_2s_3.
\end{equation}
We compare this volume with the QSE entanglement bound $V_\star=4\pi/81$ that can be derived from Eq.~\eqref{eq:ellipse_ppt}; an ellipsoid with $V_{\rm ell}>V_\star$ certifies entanglement~\cite{Jevtic:2014icx}.

\vspace{0.5em}

\noindent\textbf{QSE surface area: $A_{\rm ell}$.}

The surface area, denoted $A_{\rm ell}$, is calculated numerically from the semi-axes $s_i$, since a generic triaxial ellipsoid has no elementary closed form.

\vspace{0.5em}

\noindent\textbf{QSE orientation: $\theta_{23}$.}

In the helicity basis relevant for $\ttbar$ production, the nontrivial orientation reduces to a rotation in the scattering plane, which we take to be the $2$-$3$ plane. We quantify this rotation by
\begin{equation}
    \label{eq:tan2theta_rk}
    \tan(2\theta_{\rm 23})=\frac{2Q_{\rm 23}}{Q_{\rm 33}-Q_{\rm 22}}.
\end{equation}
Together, these observables characterize the position, size, shape, orientation, entanglement, EPR steerability, and Bell nonlocality of the reconstructed spin state. In the following sections, we compare these geometric and quantum information observables with the Fano coefficients as probes of the $\ttbar$ spin state.

\section{Reconstructing the \ttbar steering ellipsoid}
\label{sec:ttbar_qse_reconstruction}
The general construction of the QSE for the collider states described above can be realized concretely in dileptonic \ttbar production. Among SM processes, the \ttbar system is particularly suited for spin tomography because the top quark (\Pqt) decays before hadronization or significant spin decorrelation can occur~\cite{CDF:1995wbb, D0:1995jca, ParticleDataGroup:2024cfk}. The spin information prepared in the hard scattering is therefore retained in the angular distributions of its decay products~\cite{CMS:2019nrx, Mahlon:2010gw, Bernreuther2015, Han:2023fci}. The charged lepton decay products provide nearly optimal spin analyzers, allowing the conditional polarization vectors of one top spin to be reconstructed by binning the analyzer direction of the other. 

The experimental accessibility of this spin information has been demonstrated by full spin-state tomographies performed by the CMS Collaboration~\cite{CMS:2019nrx, CMS:2024zkc}, as well as by observing entanglement in the \ttbar spin system by both the ATLAS and CMS Collaborations~\cite{ATLAS:2023fsd, CMS:2024pts, CMS:2024zkc}. These results establish the spin-density matrix as an experimentally accessible object in \ttbar production, and motivate the use of the same spin-analyzer information to reconstruct its associated steering ellipsoid.\\

Having formulated the QSE as a family of collider observables, we now demonstrate their reconstruction in \ttbar production. Using NLO simulated events with spin-correlated decays, we reconstruct the steering ellipsoid from binned conditional asymmetries and compare the resulting geometry with an independent LO prediction of the underlying spin density matrix. This establishes the Standard Model benchmark for the phenomenological studies that follow.

\subsection{The \ttbar spin system}
\begin{figure}[h]
    \centering
    \begin{tikzpicture}[
    >=Latex,
    scale=1.2,
    every node/.style={font=\large},
    axis/.style={black, dashed, line width=0.8pt},
    mom/.style={black, -{Latex[length=2.5mm]}, line width=0.8pt},
    kred/.style={red!85!black, -{Latex[length=3mm]}, line width=1.1pt},
    rblue/.style={blue!80!cyan, -{Latex[length=3mm]}, line width=1.1pt},
    tdash/.style={red!85!black, dashed, -{Latex[length=3mm]}, line width=1.1pt},
    nnode/.style={circle, draw=cyan!70!black, fill=white, line width=0.8pt, minimum size=3mm, inner sep=0pt},
    ninner/.style={circle, draw=cyan!70!black, fill=cyan, line width=0.8pt, minimum size=1mm, inner sep=0pt},
]

\draw[axis] (-3.2,0) -- (3.2,0);

\draw[mom] (-1.3,0) -- (-0.7,0);
\node at (-1.55,-0.22) {$\rm p$};

\draw[mom] (1.3,0) -- (0.7,0);
\node at (1.65,-0.22) {$\rm p$};

\node[nnode] (O) at (0,0) {};
\node[cyan!70!black] at (0,-0.32) {$\nhat$};

\node[ninner] (Oin) at (0,0){};

\draw[kred] (O) -- (1.95,1.95);
\node[red!85!black] at (0.55,1.25) {$\khat$};

\draw[rblue] (O) -- (2.05,-1.70);
\node[blue!80!cyan] at (1.45,-0.95) {$\rhat$};

\draw[tdash] (O) -- (-1.85,-1.70);
\node[red!85!black] at (-1.75,-0.95) {$-\khat$};

\draw[black, line width=0.8pt] (0.55,0) arc[start angle=0,end angle=45,radius=0.55];
\node at (0.85,0.35) {$\theta^*$};

\end{tikzpicture}
    \caption{Helicity basis used for the \ttbar spin system in \pp collisions. The signs of $\rhat$ and $\nhat$ flip at $\theta^*=\pi/2$, as shown in Eqs.~\eqref{eq:nhat definition} and~\eqref{eq:rhat definition}.}
    \label{fig:helicity basis}
\end{figure}
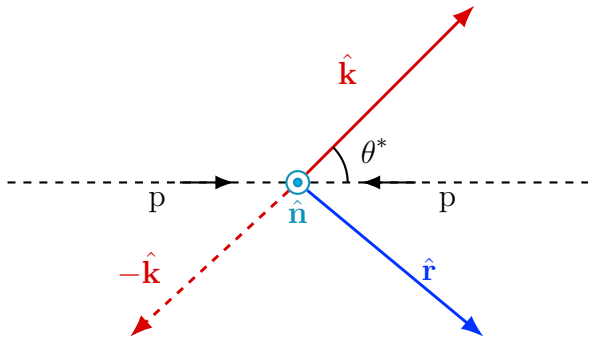

We express the \ttbar spin density matrix in an event-by-event basis known as the helicity basis~\footnote{This type of basis reconstructs properties of the so-called ``fictitious'' state~\cite{Cheng:2023qmz}, which can be related to the underlying quantum state for convex observables.} $\{\nhat,\rhat,\khat\}$~\cite{Bernreuther2015, CMS:2019nrx}, shown in Fig.~\ref{fig:helicity basis}. In the zero-momentum frame of the \ttbar system, $\khat$ is chosen along the momentum of the top quark, $\rhat$ lies in the production plane, and $\nhat$ is normal to that plane. Explicitly,
\begin{align}
    \nhat &=\operatorname{sgn}(\cos{\theta^*})\; \frac{\hat{\mathbf p} \times \khat}{\sin{\theta^*}},\label{eq:nhat definition}\\
    \rhat &=\operatorname{sgn}(\cos{\theta^*})\; \frac{\hat{\mathbf p} - \khat \cos \theta^*}{\sin{\theta^*}},\label{eq:rhat definition}
\end{align}
where $\hat{\vb p}$ is the direction of the incident proton beam and $\theta^*$ is the scattering angle between $\hat{\vb p}$ and $\khat$. The sign factor enforces the symmetry of the proton-proton (\pp) initial state under beam interchange~\cite{Bernreuther2015}. We use the same spin-quantization axes for the top quark and antiquark, following Refs.~\cite{Aguilar-Saavedra:2022uye, Aoude:2025jzc}.~\footnote{This convention differs from the original convention of Ref.~\cite{Bernreuther2015} and induces the corresponding sign change in the spin correlation and top antiquark polarization coefficients.}

In the Standard Model, inclusive \ttbar production yields mixed spin states with sizable correlations and small polarizations~\cite{Bernreuther2015, Afik:2020onf}. 
In this basis, parity-violating interactions can generate polarizations along the $k$ and $r$ axes, while normal polarizations and correlations involving the $n$ axis arise from absorptive contributions~\cite{Bernreuther2015}. Interactions that violate the combined charge-conjugation and parity (CP) symmetry can also generate asymmetries between off-diagonal spin correlation coefficients. 

We focus on the dileptonic final state, in which both \PW bosons from the top quark and antiquark decay leptonically. The charged leptons are especially clean spin analyzers, with analyzing powers close to unity, which makes this channel well suited not only for spin correlation measurement but also for the QSE-based geometric observable framework developed in this work~\cite{CMS-PAS-FTR-18-034, CMS:2019nrx, CMS:2024pts, ATLAS:2023fsd}. We identify subsystem $\text{A}$ with the top quark and subsystem $\text{B}$ with the top antiquark, although the construction can be repeated with the roles interchanged. 

We use the charged lepton from the top antiquark decay as Bob's analyzer. Conditioning on its direction $\OmegaB$ corresponds to conditioning on a local spin-analyzer POVM outcome for subsystem $\text B$. As this conditioning direction is varied, the corresponding conditional polarization vectors of the top quark, $\bm B_{\text A}(\bm m')$, approximately trace the surface of the top quark QSE.

\subsection{Computational setup}
\label{subsec:comp_setup}

We use two complementary descriptions of \ttbar production. The first is an event-level description based on Monte Carlo (MC) simulated dileptonic $\ttbar$ events, with the finite angular binning treated as described in Section~\ref{subsec:coarse-grained-steering}. Proton-proton collisions are generated with a center-of-mass energy $\sqrt{s}=13.6\,\text{TeV}$. The hard process is generated at NLO in the QCD coupling using \Powhegvtwo ~\cite{Alioli:2010xd, Frixione:2007vw}, while parton showering is performed with \Pythia~\cite{Sjostrand:2014zea}. The sample is normalized to the next-to-next-to-leading-order (NNLO) prediction for the inclusive \ttbar cross section at $\sqrt{s}=13.6\,\text{TeV}$, including next-to-next-to-leading-logarithmic (NNLL) soft-gluon resummation~\cite{Czakon:2012pz,Czakon:2011xx}. We consider integrated luminosities, $\mathcal{L}$, of $300\,\mathrm{fb}^{-1}$ and $3000\,\mathrm{fb}^{-1}$, representative of the full LHC $\pp$ dataset at $\sqrt{s}=13.6\,\text{TeV}$ and the expected High-Luminosity LHC (HL-LHC) $\pp$ dataset, respectively. For additional details on the generator parameters, we refer to Ref.~\cite{CMS:2023qyl}. We employ the \texttt{NNPDF\_nlo\_as\_0118} NLO PDF set~\cite{Ball:2017nwa}, with the mass of the top quark set to $\mtop=172.5\,\text{GeV}$, and evaluate the event kinematics before modeling detector resolution and acceptance. The directions of the charged lepton decay products are then used to reconstruct the binned conditional polarization vectors.

The second description is an independent LO QCD analytic prediction for the $\ttbar$ production spin matrix. Using the amplitude-level construction in Eq.~\eqref{eq: R matrix from scattering}, we compute the partonic $\qqbar\to\ttbar$ and $\gluglu\to\ttbar$ contributions as functions of the top-quark velocity ${\beta=\sqrt{1-4m_t^2/\hat s}}$ and scattering angle $z=\cos\theta^*$. The corresponding parton luminosities weight these partonic spin matrices to obtain the proton-proton production matrix,
\begin{equation}
R_{\rm SM}(\beta,z)
=
\sum_{I=\qqbar,\gluglu}
L_I(\beta)\,R_I(\beta,z),
\end{equation}
where $I$ labels the initial partonic channel. Inclusive or phase space-restricted observables are obtained by integrating this matrix over the corresponding $(\beta,z)$ region before extracting the normalized Fano coefficients. The Fano coefficients and QSE parameters are then obtained from Eqs.~\eqref{eq:Quantum State From R}, \eqref{eq:fanoQSECent}, and~\eqref{eq:transformationDef}. This LO QCD calculation provides an independent SM reference for validating the event-level reconstruction. 

\subsection{Reconstruction and validation}
Using the simulated events described above, we reconstruct the QSE from the finite-bin conditioning map developed in Section~\ref{subsec:coarse-grained-steering}. For each bin $\Delta\Omega_{\text B}^{(j)}$ in Bob's analyzer direction, the effective POVM Bloch vector is $\bm m'_j$ as defined in Eq.~\eqref{eq:binned_povm_element}.

\begin{figure*}[t!]
    \centering
    \includegraphics[width=0.9\linewidth]{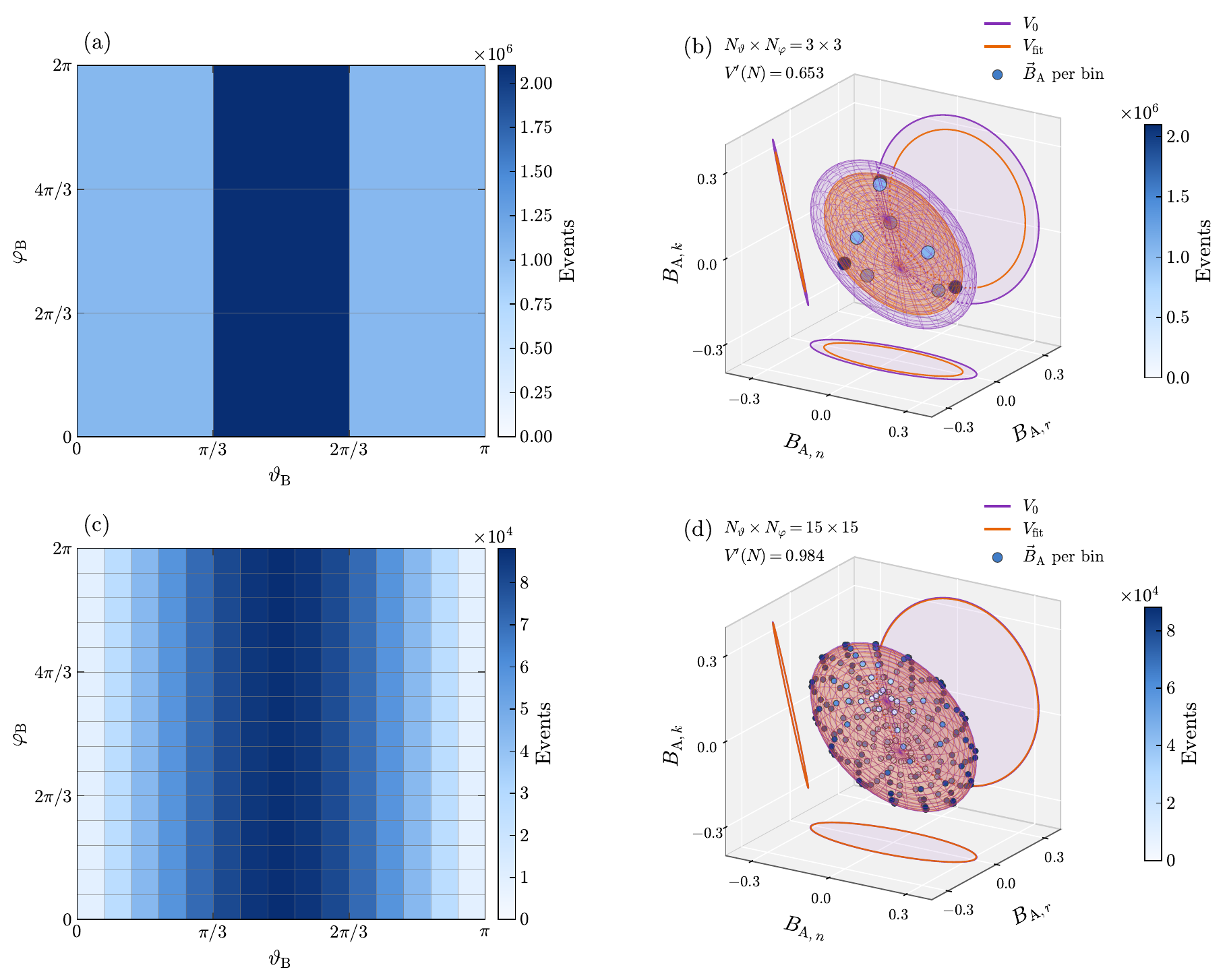}
    \caption{
    Event populations and reconstructed QSE points for two angular partitions. Panels~(a) and~(b) show the coarse $N_\vartheta=3$, $N_\varphi=3$ partition, while panels~(c) and~(d) show a more fine $N_\vartheta=15$, $N_\varphi=15$ partition. The yellow line shows the fitted ellipsoid which is significantly smaller than the predicted volume $V^0$ in the coarse partition while it almost coincide with the $V^0$ for the finer partition.}
    \label{fig:coarse_graining_scatter}
\end{figure*}

The corresponding conditional polarization vector of subsystem $\text A$ is extracted component by component from single-differential asymmetries. For $\alpha\in\{n,r,k\}$,
\begin{align}
\label{eq:bin-asym}
B_{\text A,\alpha}^{(j)}
&=
\frac{2}{\kappa_{\text A}}
\frac{
N_{\alpha,+}^{(j)} - N_{\alpha,-}^{(j)}
}{
N_j
}, \\
N_j
&\equiv
N\!\left(\OmegaB\in\Delta\Omega_{\text B}^{(j)}\right),
\nonumber \\
N_{\alpha,\pm}^{(j)}
&\equiv
N\!\left(
\OmegaB\in\Delta\Omega_{\text B}^{(j)},\,
\pm\cos\vartheta_{\text A,\alpha}>0
\right).
\nonumber
\end{align}
Here $N_j$ and $N_{\alpha,\pm}^{(j)}$ are the event counts, and $\vartheta_{\text A,\alpha}$ is the polar angle of Alice's analyzer with respect to the $\alpha$ axis. Figure~\ref{fig:coarse_graining_scatter} shows the bin population and the reconstructed QSE volumes for two different angular partitions: $N=15$ and $N=300$. We evaluate the event counts and analyzer angles before modeling detector resolution and acceptance. In an experimental analysis, the same asymmetries can be formed from angular distributions corrected for detector resolution, acceptance, and event-selection effects. The reconstructed vectors
\begin{equation} 
\vb x_j 
\equiv \BA^{(j)} 
= \BA(\bm m'_j), \qquad j=1,\ldots,N_{\rm bin}, 
\end{equation} 
provide a coarse-grained sampling of the $\Pqt$-quark QSE.

To fit the ellipsoid, we use the marginal and conditional distributions from Eq.~\eqref{eq:twofold_angular_xsec} and~\eqref{eq:conditional_angular_distribution}, and the binned counts in Eq.~\eqref{eq:bin-asym} to perform a maximum-likelihood fit based on the predicted bin probabilities. Integrating the conditional distribution of Eq.~\eqref{eq:conditional_angular_distribution} over the hemisphere, we obtain the probability for $\cos\vartheta_{\text{A},\alpha}>0$
\begin{equation}
  p^{(j)}_{\alpha}
  = \frac{1}{2}\left(1 + \frac{1}{2}\kappa_{\text{A}}B_{{\rm A},\alpha}^{\,(j)}\!\right),
  \label{eq:hemiprob}
\end{equation}
where $\vec{B}_{\text{A}}^{\,(j)}$, determined by Eq.~\eqref{eq:steering_map_collider}, is the steering map from the unit sphere vector to the QSE and is a function of the Fano coefficients. Integrating the marginal distribution of
Eq.~\eqref{eq:twofold_angular_xsec} over the solid angle patch gives
\begin{equation}
  P_j = \frac{\Delta_j}{4\pi}
        \left(1 + \BB\!\cdot\!\bm m'_j\right).
  \label{eq:margprob}
\end{equation}
The likelihood is then the corresponding product of the binomial conditional and the marginal multinomial 
\begin{align}
  \ln\mathcal{L}
  &= \sum_{j}\sum_{\alpha}
     \left[\, N^{(j)}_{\alpha,+} \ln p^{(j)}_{\alpha}
            + N^{(j)}_{\alpha,-} \ln\!\left(1 - p^{(j)}_{\alpha}\right)
     \right]
  \nonumber\\
  &\quad + \sum_{j} N_j \ln P_j.
  \label{eq:likelihood}
\end{align}
The marginal factor constrains Bob's polarization, while each conditional  factor directly probes the steering-map image $\vec{B}_{\text{A}}^{\,(j)}$ of
Eq.~\eqref{eq:steering_map_collider}. The likelihood can be maximized over the fifteen normalized Fano coefficients $(\BA, \BB, C)$, from which the QSE center and shape matrix can be determined.

Equivalently, the fit can be performed directly in the steering
parameterization $(\cA,\, \TA,\, \BB)$. This parameterization carries the same fifteen degrees of freedom, and the Fano coefficients are recovered through the inverse map
\begin{align}
  \BA &= \cA + \TA\,\BB ,
  \nonumber\\
  C &= \cA\,\BB^{\top}
     + \frac{\TA}{\gammaB} 
     + \frac{\gammaB - 1}{\gammaB\, \|\BB\|^{2}}\,
       \TA\,\BB \BB^{\top} .
  \label{eq:canonicalmap}
\end{align}
Fitting in this basis yields the geometric QSE observables and their uncertainties directly, without nonlinear propagation from the Fano coefficients.

Because the probabilities in Eq.~\eqref{eq:hemiprob} and~\eqref{eq:margprob} are exact solid-angle integrals of the predicted probability density over each patch, the model retrieves the ellipsoid without the coarse-grain bias introduced by the construction. Refinement or coarsening the partition changes only the statistical resolution of the fit, and the reconstructed ellipsoid converges to the underlying QSE independently of the bin size.

\begin{figure}[t]
    \centering
    \includegraphics[width=\linewidth]{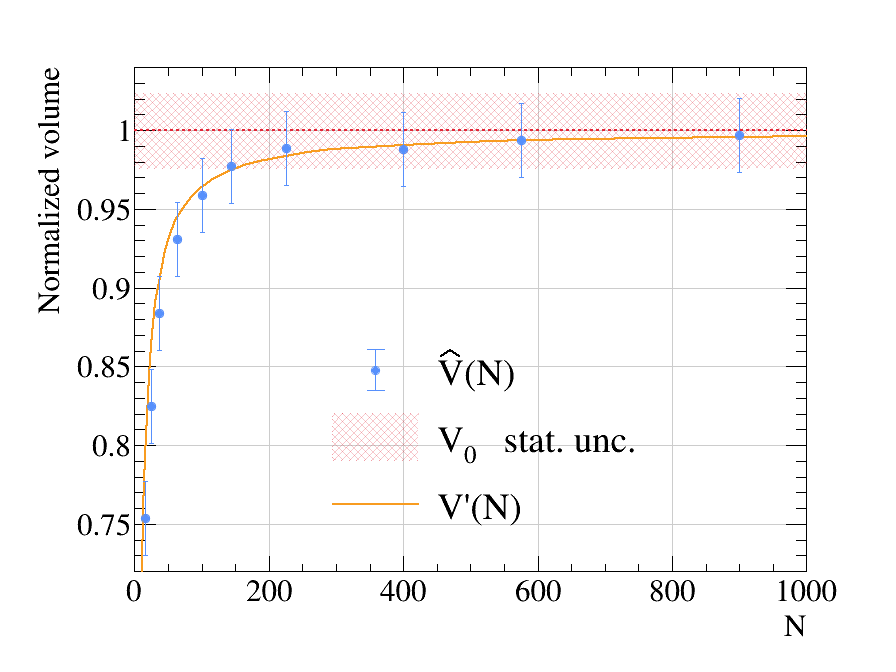}
    \caption{Normalized volume $\widehat V=V_{\rm fit}/V_0$ (blue circles) reconstructed from binned conditional polarization vectors as a function of the total number of angular bins $N_{\rm bin}=N_\vartheta N_\varphi$, where $N_\varphi=N_\vartheta=\sqrt{N}$ with the bin-center prescription. The coarse-grained prediction $V'(N)$ (orange line) from Eq.~\eqref{eq:coarse_grained_volume_factor} is shown for comparison. The horizontal line (red dashed line) denotes the tomography reference, and the shaded band shows the statistical uncertainty from the finite size of the simulated sample.}
    \label{fig:Volume_coarse}
\end{figure}

\begin{figure}[t]
    \centering
    \includegraphics[width=\linewidth]{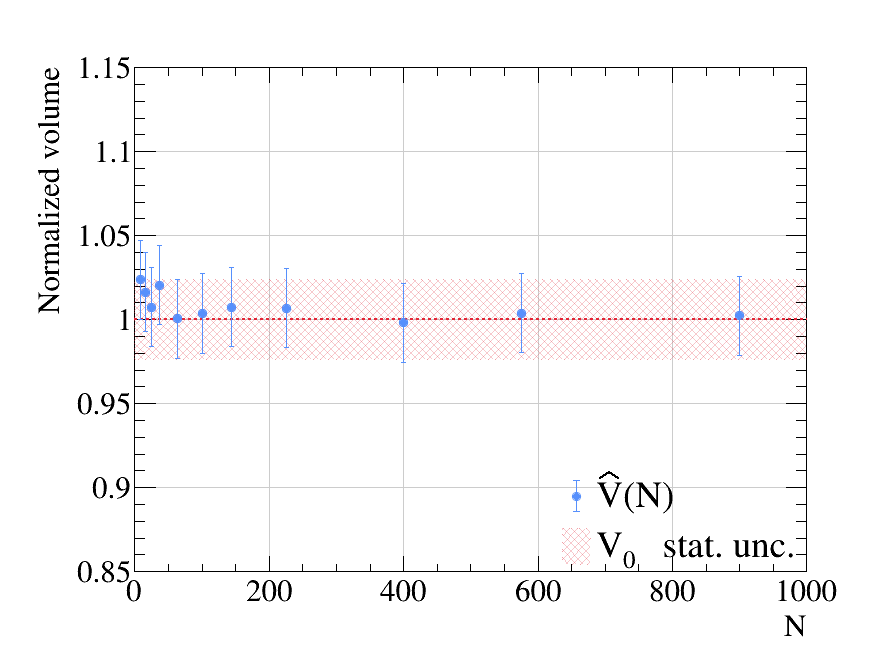}
    \caption{Normalized volume $\widehat V=V_{\rm fit}/V_0$ (blue circles) reconstructed from binned conditional polarization vectors as a function of the total number of angular bins $N_{\rm bin}=N_\vartheta N_\varphi$, where $N_\varphi=N_\vartheta=\sqrt{N}$, with the patch-integrated likelihood. The horizontal line (red dashed line) denotes the tomography reference, and the shaded band shows the statistical uncertainty from the finite size of the simulated sample.}
    \label{fig:Volume}
\end{figure}

The coarse-grain picture of Section~\ref{subsec:coarse-grained-steering} can be recovered if the patch-averaged direction $\overline{\Omega}_B^{(j)}$ in Eq.~\eqref{eq:margprob} and in the steering map of Eq.~\eqref {eq:steering_map_collider} is replaced by the unit bin-center direction. Each conditioning bin is treated as an ideal measurement along its central axis. The mismatch between this assumed unit direction and the true averaged direction,
$\|\overline{\Omega}_B^{(j)}\| < 1$, contracts the fitted correlation
matrix exactly as one would get by fitting a coarse-grained steering ellipsoid from its points, and the reconstructed volume acquires the suppression $V'(N_\vartheta, N_\varphi)$ of Eq.~\eqref{eq:coarse_grained_ellipsoid_volume}.
Figures~\ref{fig:Volume_coarse} and~\ref{fig:Volume} demonstrate this correspondence: the volume reconstructed with the bin-center prescription tracks $V'(N_\vartheta, N_\varphi)$ and approaches the tomography reference $V_0$ only as the partition is refined, whereas the patch-integrated likelihood reproduces the reference volume at any partition. The residual disagreement between the fitted volume and the scale factor prediction for the coarse partitions at low $N$ is expected, since large bins do not produce an exactly uniform contraction of the steering ellipsoid.
To quantify the effect of finite angular binning, we compare the fitted ellipsoid volume, $V_{\rm fit}(N_\vartheta,\, N_\varphi)$, with a tomography reference volume $V_0$. The reference is obtained by extracting $\BA$, $\BB$, and $C$ directly from the forward-backward asymmetries of the complete angular distributions~\cite{Bernreuther2015}.

As a final validation, we compare the QSE and quantum information observables obtained from the reconstructed event sample with those obtained from the independent LO analytic calculation. Both calculations are integrated over the full phase space, and the results are shown in Table~\ref{tab:qit_observables_qcd}. The comparison is not expected to be exact because the event sample is generated at NLO, while the analytic reference is evaluated at LO. Nevertheless, the two calculations give compatible SM reference values when compared with the fixed-order calculations in Ref.~\cite{Behring:2019iiv} for the observables used below. The quoted theory uncertainties are obtained from enveloping the seven-point renormalization and factorization scale variations, together with the PDF and parton shower variations used for the simulated sample.

\begin{table}[!ht]
    \caption{Standard Model QCD reference values for the QSE and QIS observables used in the analysis. The analytic column is obtained from the independent LO production spin matrix, while the MC column is obtained from the reconstructed NLO simulated event sample before detector modeling. The quoted analytic uncertainties are obtained by enveloping the seven-point renormalization and factorization scale variations as well as PDF uncertainties. The MC-quoted uncertainties also include parton-showering variations.}
    \label{tab:qit_observables_qcd}
    \centering
    \renewcommand{\arraystretch}{1.5}
    \setlength{\tabcolsep}{10pt}

    \begin{tabular}{ccc}
        \textbf{Observable} &
        \textbf{Analytic} &
        \textbf{MC} \\
        \hline
        $\xi$
        & $\phantom{-}0.4962^{+0.0015}_{-0.0029}$
        & $\phantom{-}0.558 \pm 0.017$ \\

        $A_{\rm ell}$
        & $\phantom{-}0.7537^{+0.0019}_{-0.0043}$
        & $\phantom{-}0.690 \pm0.017$ \\

        $V_{\rm ell}$
        & $\phantom{-}0.0212^{+0.0013}_{-0.0009}$
        & $\phantom{-}0.004 \pm 0.005$ \\

        $\tan(2\theta_{rk})$
        & $\phantom{-}0.7211^{+0.0080}_{-0.0114}$
        & $\phantom{-}0.768 \pm0.015$ \\

        $\Delta_+$
        & $-1.0020^{+0.0047}_{-0.0061}$
        & $-0.964 \pm 0.016$ \\

        $\Delta_-$
        & $-0.3471^{+0.0001}_{-0.0005}$
        & $-0.419 \pm 0.004$ \\

        $\hat{S}$
        & $\phantom{-}0.4657^{+0.0003}_{-0.0010}$
        & $\phantom{-}0.518 \pm 0.006$ \\

        $\mathcal{I}_{\rm CHSH}$
        & $-0.7705^{+0.0004}_{-0.0009}$
        & $-0.783 \pm 0.004$ \\
    \end{tabular}
\end{table}

Having validated the reconstruction and fixed the inclusive SM reference geometry, we next apply the same QSE observables in two settings: first to SM \ttbar production in the boosted central region, where the inferred spin density matrix is expected to lie in the CHSH-violating region, and then to SMEFT deformations of the \ttbar spin state.

\section{Steering geometry and Bell nonlocality in the Standard Model}
\label{sec:qse_bell_ineq}

The inclusive SM \ttbar\ state provides the reference geometry for the reconstructed steering ellipsoid. To benchmark our QSE-based geometric observable framework in a regime of enhanced quantum correlations, we additionally consider the boosted central region, where previous studies have shown that entanglement, steering, and Bell sensitivity are maximized~\cite{Afik:2020onf, Afik:2022dgh, Aguilar-Saavedra:2022wam}. We adopt the phase space selection,
\begin{equation}\label{eq:Boosted central region}
    \beta\geq 0.94,
    \qquad
    |z|<0.2,
\end{equation}
which matches the phase space region defined in Ref.~\cite{Aguilar-Saavedra:2022uye}. In this region, the \ttbar spin state approaches the Bell-state configuration~\cite{Aoude:2022imd, CMS:2025brx}
\begin{equation}
\label{eq:bell_triplet}
    \ket{\Psi^+}
    =
    \frac{1}{\sqrt{2}}
    \left(
        \ket{\uparrow\downarrow}
        +\ket{\downarrow\uparrow}
    \right).
\end{equation}
In this phase-space region, the polarizations are small, so the QSE shape matrix is well approximated by $Q\simeq CC^\top$. As discussed in Section~\ref{subsec:qis_qse_observables}, the two longest QSE semi-axes then give the geometric realization of the Horodecki CHSH criterion (see Eq.~\eqref{eq:qse_bell_ineq}). The boosted central region therefore provides a Standard Model setting in which to compare the conventional Fano-basis Bell observable (see Eq.~\eqref{eq:bell_cii_cjj}) with its QSE-based geometric counterpart.

The Bell observables considered here characterize the reconstructed \ttbar\ spin density matrix rather than constituting a direct Bell test. Experimentally, the measured quantities are the final-state momenta, from which the spin density matrix is reconstructed using the SM spin-analyzing powers encoded in the corresponding POVM elements. The Bell indicators in Eqs.~\eqref{eq:bell_cii_cjj} and~\eqref{eq:qse_bell_ineq} should therefore be interpreted as reconstructed expectation values of Bell operators for the inferred spin density matrix, rather than as a Bell test of local realism. This distinction follows the original observation of Ref.~\cite{ABEL1992304} and subsequent discussions of Bell inequalities constructed from commuting collider observables~\cite{Bechtle:2025ugc, Abel:2025wwa}. Consequently, a positive Bell indicator implies that the reconstructed \ttbar\ spin state lies in the CHSH-violating region, but does not by itself constitute a Bell test capable of excluding local-hidden-variable descriptions of the measured collider observables.

Within the boosted central region, the spin-triplet state of Eq.~\eqref{eq:bell_triplet} has the diagonal spin correlation matrix $C=\mathrm{Diag}(-1,\,+1,\,+1)$. Consequently, the CHSH-sensitive directions correspond to the $(n,r)$ or $(r,k)$ axes~\cite{Aguilar-Saavedra:2022uye}. We find that the $(n,r)$ axes provide the strongest sensitivity and therefore use
\begin{equation}
    \label{eq:chsh_obs}
    \mathcal I_2
    \equiv
    \mathcal I_{rn}^{(-)}
    =
    |\Crr-\Cnn|-\sqrt{2}.
\end{equation}
We compare this conventional Fano-based Bell observable with its QSE-based geometric counterparts, $\mathcal I_{\rm QSE}^{(1)}$ and $\mathcal I_{\rm QSE}^{(2)}$, introduced in Section~\ref{subsec:qis_qse_observables}. We estimate these observables' values and uncertainties using the simulated sample with the same settings described at the beginning of Section~\ref{subsec:comp_setup}. We assume that 12\% of events are detected and selected for analysis, which is a reasonable assumption for a dilepton \ttbar analysis~\cite{CMS:2019nrx}. We estimate statistical uncertainties and correlations from $10\,000$ bootstrap replicas, constructed by Poisson resampling the simulated events. The set of systematic uncertainties considered is the same as those in Table~\ref{tab:qit_observables_qcd}.

Table~\ref{tab:bell} summarizes the projected values and significances of the indicators considered. We also report the correlated combination of the conventional Fano indicator $\mathcal I_2$ and the QSE-based indicator $\mathcal I_{\rm QSE}^{(2)}$, constructed from their two-dimensional bootstrap covariance matrix with an estimated correlation of $\rho(\mathcal I_2,\mathcal I_{\rm QSE}^{(2)})\simeq0.92$. The resulting significance shows that a complementary set of observables provides access to the Bell indicators at the same significance level. This clearly demonstrates the value of this new approach.

\begin{table}
\caption{Projected CHSH indicators in the boosted central $t\bar t$ region for dileptonic final states. The first number indicates the statistical uncertainty while the second refers to the systematic uncertainty.}
\label{tab:bell}
\begin{tabular}{lcc}
Luminosity & $300\,\mathrm{fb}^{-1}$ & $3000\,\mathrm{fb}^{-1}$ \\
\hline
$\mathcal I_2$                     & $0.120 \pm 0.135 \pm 0.019$ & $0.120 \pm 0.043 \pm 0.013$ \Tstrut \\
$\mathcal I_{\rm QSE}^{(2)}$       & $0.123 \pm 0.128 \pm 0.017$  & $0.123 \pm 0.043 \pm 0.012$ 
\\[1.5ex]
$Z(\mathcal I_2)$                  & $0.88$ & $2.67$ \\ 
$Z(\mathcal I_{\rm QSE}^{(2)})$    & $0.95$ & $2.79$ \\
$Z_{\rm comb.}$                    & $0.97$ & $2.80$ \\
\end{tabular}
\end{table}

\section{New-physics deformations of \ttbar steering ellipsoids}
\label{sec:smeft_qse_deformations} 

Having established the collider reconstruction of QSE-based geometric observables, we now investigate how heavy new physics deforms the reconstructed quantum-state geometry within the Standard Model Effective Field Theory (SMEFT). In this framework, higher-dimensional interactions modify the $\ttbar$ production amplitudes and thereby deform both the reconstructed Fano coefficients $\{\BA,\,\BB,\, C\}$ and the associated quantum steering ellipsoid. SMEFT effects on \ttbar spin correlations and entanglement have been studied in Refs.~\cite{Aoude:2022imd, Severi:2022qjy}, while more recent work has explored additional quantum information probes of new physics in the top quark sector~\cite{Aoude:2025jzc, Subba:2026nzs}. Here we extend these studies by examining how the same SMEFT deformations are encoded in the geometric observables of the quantum steering ellipsoid.

We describe the new interactions within the SMEFT framework using a slightly modified Warsaw basis~\cite{Grzadkowski:2010es, Aoude:2022imd}
\begin{equation}
    \mathcal{L}_{\text{SMEFT}} = \mathcal{L}_{\text{SM}} + \frac{1}{\Lambda^2}\sum_i c_i \mathcal{O}_i\,,
\end{equation} 
where $\Lambda$ is the characteristic scale of heavy new physics and the Wilson coefficients $c_i$ parameterize the strengths of the dimension-six operators $\mathcal O_i$. We restricted the analysis to $\text{CP}$-even operators. The operators separate naturally by production channel: $\mathcal O_G$, $\mathcal O_{\phi G}$, and $\OtG$ modify the \gluglu contribution. In contrast, the color-octet and color-singlet four-fermion operators modify the \qqbar contribution. The coefficient values shown in the figures later in this Section are chosen to display the qualitative geometric response.

The SMEFT dependence is included by extending the LO QCD analytic spin density construction of Section~\ref{subsec:comp_setup}. At fixed partonic kinematics, described by the top quark velocity $\beta$ and scattering angle variable $z$, the production spin density matrix is written as
\begin{align}\label{eq: R expansion in terms of Lambda}
R(\beta,z;c_i)
=
R_{\rm SM}(\beta,z)
&+
\frac{1}{\Lambda^2}
\sum_i c_i R_i^{\rm int}(\beta,z)\\
&+
\frac{1}{\Lambda^4}
\sum_{i,j} c_i c_j R_{ij}^{\rm quad}(\beta,z).\nonumber
\end{align}
Here, $R_i^{\rm int}$ denotes the interference between the SM amplitude and the operator $\mathcal O_i$, while $R_{ij}^{\rm quad}$ denotes the contribution quadratic in dimension-six amplitudes. Each term is expanded in the Fano basis of Eq.~\eqref{eq:Spin density matrix}, with the corresponding parton-level results listed in Appendix~\ref{app: Fano Coefficients DEF}. Convolution with the parton luminosities and integration over the chosen phase space region then determine the Wilson-coefficient dependence of both the Fano coefficients and the associated QSE observables.

\subsection{Response to effective interactions}

\begin{figure*}[t!]
    \centering
    \includegraphics[width=\linewidth]{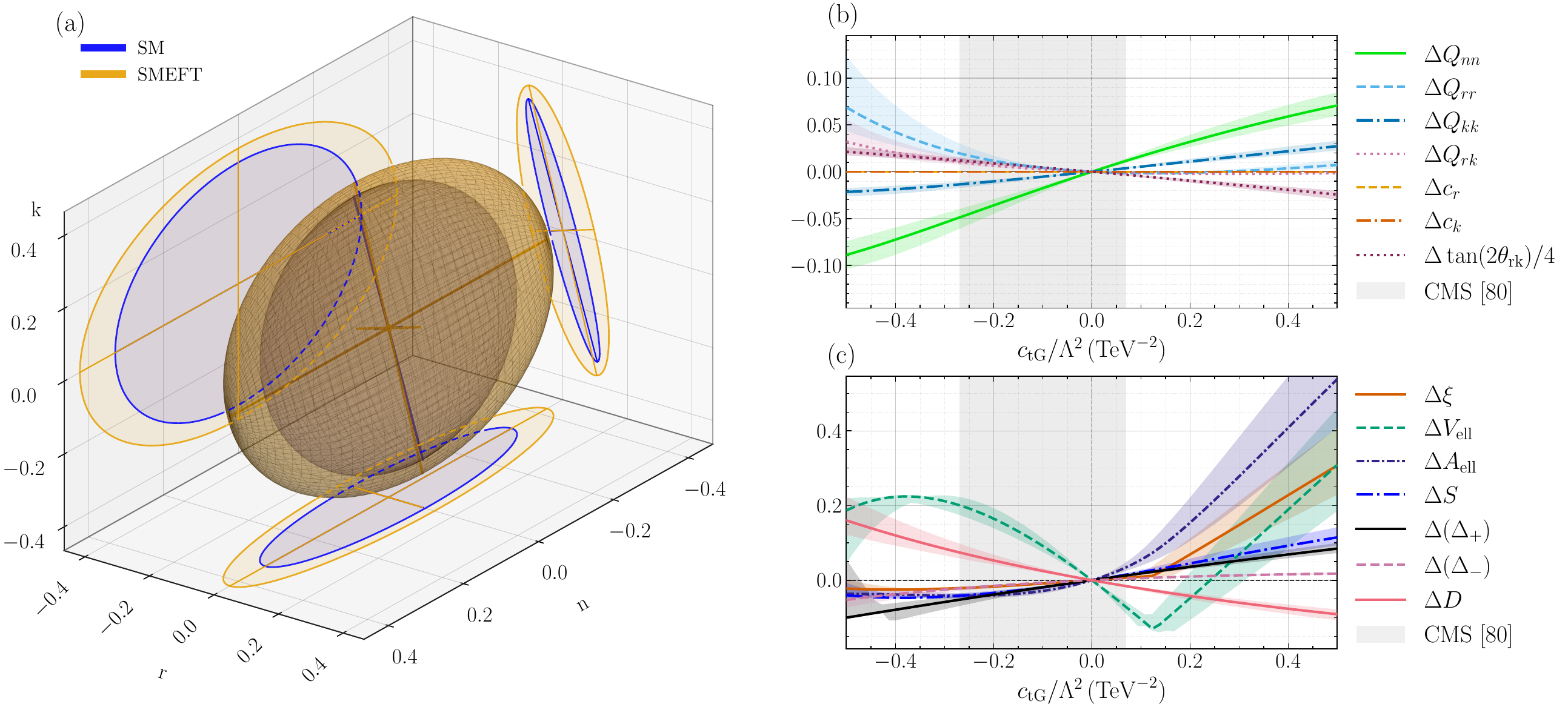}
    \caption{Inclusive QSE deformation from $\OtG$. (a): SM and SMEFT $\left(\ctG/\Lambda^2=0.5~\mathrm{TeV}^{-2}\right)$ ellipsoids. (b): shifts in QSE parameters. (c): Shifts in relevant QIS and geometric observables, where $\Delta V_{\rm ell}=(V_{\rm EFT}-V_{\rm SM})/V_*$\footnote{The reason for normalizing by $V_*$ is that the volume shifts are an order of magnitude smaller than the rest of the QIS observables.}. Shown are also theory uncertainty bands and constraints from fits performed by the CMS~\cite{CMS:2025ugn} Collaboration. Theory uncertainty bands are obtained from scale variations, numerical integration confidence, and PDF uncertainties for \pp collisions at $\sqrt s=13.6\,\rm TeV$.}
    \label{fig:ctG_Ellipsoid_Inclusive}
\end{figure*}

\begin{figure*}[t!]
    \centering
    \includegraphics[width=\linewidth]{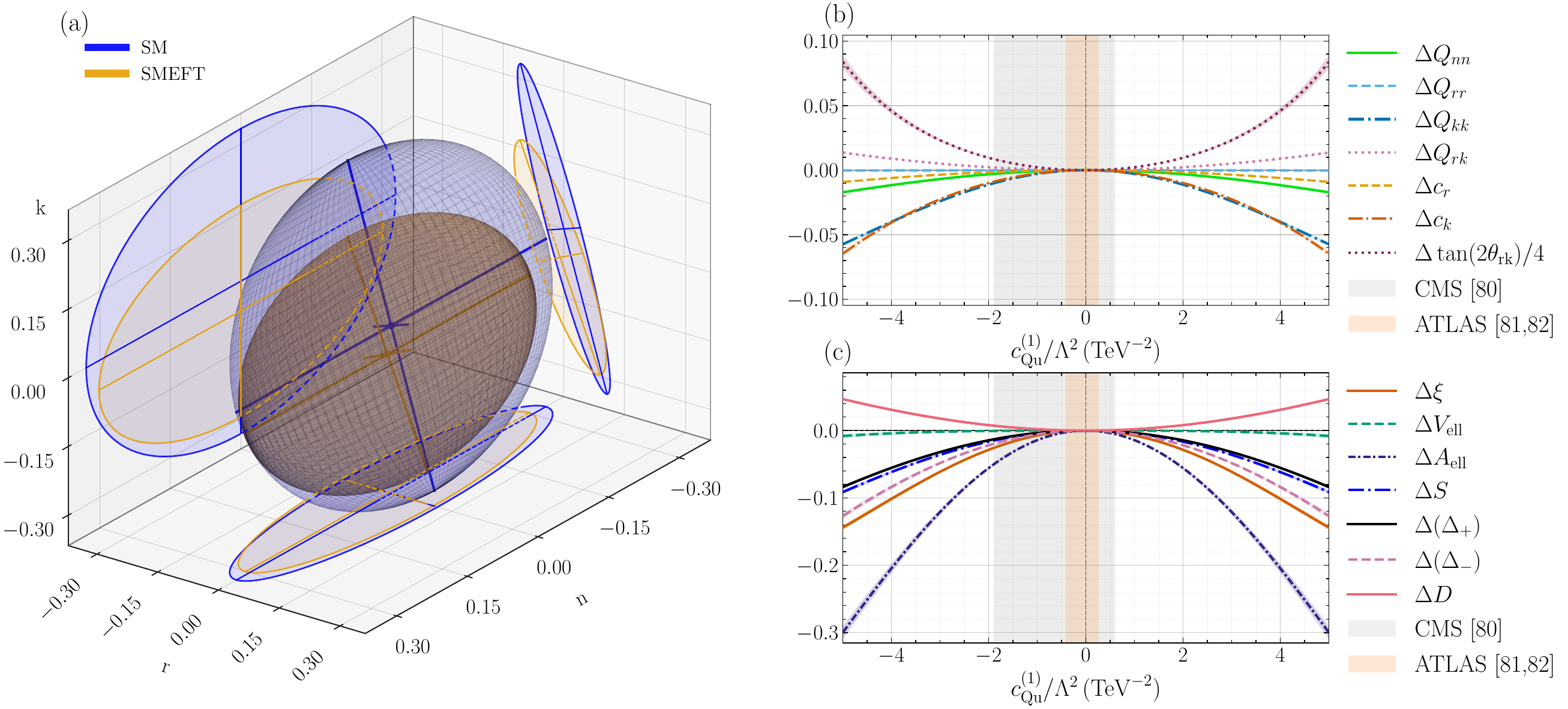}
    \caption{Inclusive QSE deformation from $\mathcal O_{\rm Qu}^{(1)}$. (a) SM and SMEFT $\left(c_{\rm Qu}^{(1)}/\Lambda^2=5\,\text{TeV}^{-2}\right)$ ellipsoids. (b): Shifts in QSE parameters. (c): Shifts in relevant QIS and geometric observables. Also shown are the fits performed by the ATLAS~\cite{ATLAS:2022waa,ATLAS:2022mlu} and CMS~\cite{CMS:2025ugn} Collaborations.}
    \label{fig:Sample 4F QSE}
\end{figure*}

We first examine how representative SMEFT interactions deform the reconstructed QSE relative to the inclusive SM geometry. Following the geometric characterization introduced in Section~\ref{subsec:qis_qse_observables}, we describe these deformations through the QSE center, semi-axes, principal directions, and derived geometric observables. The inclusive SM ellipsoid, shown in blue in Figs.~\ref{fig:ctG_Ellipsoid_Inclusive}(a) and~\ref{fig:Sample 4F QSE}(a), serves as the reference geometry.

The inclusive SM ellipsoid is centered at the origin and lies below both the EPR steerability and the CHSH-violation thresholds (see Section \ref{subsec:qis_qse_observables}). Its finite volume reflects full-rank spin correlations. However, the ellipsoid is rotated in the $r$-$k$ plane because the inclusive QCD spin correlation matrix is not diagonal in the helicity basis,
\begin{equation}
Q
=
\begin{pmatrix}
Q_{nn} & 0 & 0 \\
0 & Q_{rr} & Q_{rk} \\
0 & Q_{rk} & Q_{kk}
\end{pmatrix}.
\end{equation}
Consequently, SMEFT interactions can modify all three semi-axes, while rotations of the principal axes remain confined to the $r$-$k$ plane. The corresponding SM values are summarized in Table~\ref{tab:qit_observables_qcd} and serve as the reference geometry for the QSE and QIS deformations discussed below.

The chromomagnetic operator $\OtG$ gives a representative dipole deformation of the top quark--gluon interaction. Figure~\ref{fig:ctG_Ellipsoid_Inclusive}(a) shows the corresponding QSE deformation for a fixed value of $\ctG$. The ellipsoid center remains unchanged, whereas the size and orientation are modified, indicating that $\OtG$ primarily deforms the spin correlation structure rather than inducing net polarization.

Figure~\ref{fig:ctG_Ellipsoid_Inclusive}(b) resolves this response at the level of the QSE shape matrix. The response is anisotropic, causing the semi-axes and the $r$-$k$ rotation to evolve at different rates. Consequently, Fig.~\ref{fig:ctG_Ellipsoid_Inclusive}(c) shows that the ellipsoid volume exhibits a stronger response than several of the scalar QIS observables. The response is also asymmetric under $\ctG\to-\ctG$, reflecting the interference structure of the dipole interaction. The pronounced feature in the volume of the ellipsoid occurs near $\ctG\approx0.1325\,\textrm{TeV}^{-2}$ when one eigenvalue of $Q$ approaches zero and the ellipsoid becomes nearly degenerate. 

Four-fermion operators produce a qualitatively different geometric response. Figure~\ref{fig:Sample 4F QSE}(a) shows that the representative contact interaction both displaces and deforms the steering ellipsoid. Unlike the chromomagnetic operator, the deformation is accompanied by a translation of the ellipsoid center, reflecting the generation of nonzero single-spin polarizations while also modifying the spin correlation matrix.

Figure~\ref{fig:Sample 4F QSE}(b) resolves the deformation at the level of the QSE shape matrix. In addition to the translated center, the shape matrix is modified, changing the semi-axes and the  $r$-$k$ orientation. Figure~\ref{fig:Sample 4F QSE}(c) compares the resulting geometric observables with their QIS counterparts. The QSE therefore encodes the four-fermion response as a unified geometric deformation --- combining translation, stretching, and rotation --- rather than as separate shifts of polarization vectors and spin correlation coefficients.

\subsection{Complementarity of geometric and quantum information observables}
\label{subsec:complementarity}

The preceding SMEFT examples illustrate that the QSE observables and conventional quantum information observables provide complementary descriptions of the same reconstructed spin density matrix. 
Conventional observables such as the entanglement witnesses $\Delta_\pm$, the EPR steerability observable $\hat S$, and the Bell indicator $\mathcal I_2$ test whether the quantum state crosses specific quantum-correlation thresholds. By contrast, the QSE observables continuously characterize the geometry of the reconstructed conditional-state space through its position, principal extent, orientation, anisotropy, and size. 

This distinction is reflected in the SMEFT response. Dipole operators primarily reshape the steering ellipsoid by changing its semi-axes, orientation, and volume, whereas four-fermion operators additionally translate its center through induced single-spin polarizations. These geometric deformations do not need to track the conventional QIS observables, since they redistribute spin correlations without necessarily changing the entanglement, steering, or Bell classification of the reconstructed state. In the coefficient ranges considered here, for example, the volume of the ellipsoid exhibits a strong response while remaining below the QSE entanglement-volume threshold.

The QSE observables are therefore not redundant with conventional quantum information observables. They quantify how its conditional-state geometry evolves under heavy new physics described within the SMEFT framework. Together, conventional and geometric observables provide complementary descriptions of the same reconstructed quantum state.

\subsection{Sensitivity gains to Standard Model deformations}
\label{subsec:smeft_qse_sensitivity}

The preceding discussion established that QSE observables provide a complementary geometric description of the reconstructed quantum state. We now investigate whether this parameterization can improve the projected sensitivity to SMEFT interactions.

For the two-qubit states considered here, the QSE observables are nonlinear functions of the same polarization and spin correlation coefficients entering the Fano representation. They therefore contain no additional quantum-state information, but their SMEFT response and covariance do not necessarily align with those of the individual Fano coefficients. We quantify the resulting sensitivity gains through one-coefficient $\chi^2$ profiles, first as a function of the assumed QSE precision and subsequently using inclusive projections at two different luminosities of $\lumi =300\,\rm fb^{-1}$ ($\sqrt{s} = 13.6$ TeV) and $\lumi=3000\,\rm fb^{-1}$ ($\sqrt{s} = 14$ TeV).

For a Wilson coefficient $c$, with all remaining coefficients set to zero, we define
\begin{align}
&\chi^2_{\rm obs}(c)=\Delta\bm O^{\top}(c)\,\mathbf V_{\rm obs}^{+}\,\Delta\bm O(c),\\
&\Delta\bm O(c)=\bm O(c)-\bm O(0),\nonumber
\end{align}
where $\mathbf V_{\rm obs}^{+}$ is the Moore--Penrose pseudoinverse of the covariance matrix restricted to the selected observables. The $95\%$ confidence interval is the connected region containing $c=0$ for which $\chi^2_{\rm obs}(c)\leq\chi^2_{95}$, with $\chi^2_{95}=3.841$.

The observable set consists of an operator-dependent baseline $\mathcal B$ of Fano coefficients, determined in Table 9 of Ref.~\cite{CMS:2019nrx}, and the selected QSE observables. For the gluonic operators, the baseline $\mathcal B$ is
\begin{equation}
\bm O_{\mathcal B}=\left\{C_{kk},C_{nn},C_{rk}+C_{kr},D\right\},
\end{equation}
while the four-fermion operators additionally include the non-vanishing polarization components of the top quark spin. 

Currently, the most stringent limits on the Wilson coefficients being considered come from the ATLAS~\cite{ATLAS:2022mlu, ATLAS:2022waa} and CMS~\cite{CMS:2025ugn} Collaborations. Differences in perturbative accuracy, phase space treatment, unfolding, and parton showering prevent the uncalibrated profiles from reproducing the published intervals exactly. We therefore apply one operator-dependent global calibration to the full experimental covariance information taken from Ref.\ \cite{CMS:2019nrx}. Equivalently, $\chi^2_{\rm cal}=\alpha\chi^2_{\rm obs}$, where the calibration factor $\alpha$ is fixed to the published 95\% confidence level interval and $\left[c_-^{(0)},\,c_+^{(0)}\right]$ is the published reference interval mapped to our Wilson-coefficient convention. We apply the same calibration to the baseline and baseline-plus-QSE profiles.

\subsubsection{One observable at a time}

We first add one QSE observable $\mathcal E$ at a time to the calibrated baseline, using a center-of-mass energy $\sqrt{s}=13.6\,\rm TeV$ and $\lumi=300\,\rm fb^{-1}$ for the covariance matrix. The observables considered are
\begin{align}
\bm O_{\mathcal E} \in&\left\{Q,\bm c,V_{\rm ell},A_{\rm ell},
\tan(2\theta_{rk}),\mathcal I_{\rm QSE}^{(1)},\mathcal I_{\rm QSE}^{(2)}\right\},\\
\end{align}
with the SMEFT-induced shifts defined relative to the SM: $\Delta\mathcal E=\mathcal E(c)-\mathcal E(0)$. We do not consider the missing observables, namely the geometric entanglement $\xi$ and steerability $S$, due to their highly non-trivial dependence on the ellipsoid geometry; we leave this for future studies. The improvement in the expected limit is expressed and referred to as ``gain", which is defined as the ratio of the baseline and combined interval widths, \mbox{$G \equiv ( c_+^{\mathcal B}-c_-^{\mathcal B}) / (c_+^{\mathcal{B}+\mathcal E}-c_-^{\mathcal{B}+\mathcal{E}})$}.

To vary the precision of Q without altering the information already contained in the baseline, we rescale only its conditional uncertainty given the baseline,
\begin{align}
\zeta_\mathcal E&\equiv\frac{\sigma_{\mathcal E|\mathcal B}}{\sigma_{\mathcal E|\mathcal B}^{(0)}},
&\left(\sigma_{\mathcal E|\mathcal B}^{(0)}\right)^2
=\mathbf V_{\mathcal{EE}}-\mathbf V_{\mathcal E\mathcal B}\mathbf V_{\mathcal{BB}}^{+}\mathbf V_{\mathcal{B}\mathcal{E}}.
\end{align}
The corresponding profile can be written using the Schur complement~\cite{OUELLETTE1981187} as
\begin{align}
&\chi^2_{\mathcal{B}+\mathcal{E}}(c;\zeta_{\mathcal E})
=\\&\chi^2_{\mathcal B}(c)+\frac{\left[\Delta\mathcal E(c)-\mathbf V_{\mathcal{E}\mathcal{B}}\mathbf V_{\mathcal{BB}}^{+}\Delta\bm O_{\mathcal B}(c)\right]^2}{\zeta_{\mathcal E}^2\left(\sigma_{\mathcal{E}|\mathcal{B}}^{(0)}\right)^2}.\nonumber
\end{align}
The scan therefore measures the independent information supplied by $\mathcal E$ after profiling the baseline observables. The point $\zeta_\mathcal E=1$ reproduces the nominal calibrated $\lumi=300\,\rm fb^{-1}$ hybrid covariance, while $\zeta_{\mathcal E}<1$ and $\zeta_\mathcal E>1$ represent improved and degraded conditional precision, respectively. The reference values for $\sigma_{\mathcal{E}|\mathcal{B}}^{(0)}$ are shown in Table~\ref{tab:conditional-uncertainties}. 

\begin{table}[!ht]
\centering
\caption{Nominal conditional uncertainties of the added observables after conditioning on the complete CMS-compatible Fano baseline $\mathcal{B}$. The systematic entries are the raw marginal values $\sigma_{\mathcal{E}}^{\rm syst}=\sqrt{(V_{\rm syst})_{\mathcal{E}\mathcal{E}}}$ and are not conditioned on $\mathcal{B}$. All values shown are in units of $10^{-3}$.}
\label{tab:conditional-uncertainties}
\begin{tabular*}{\columnwidth}{@{\extracolsep{\fill}}ccccc@{}}
 & \multicolumn{2}{c}{$\lumi=300\,\mathrm{fb}^{-1}$} & \multicolumn{2}{c}{$3000\,\mathrm{fb}^{-1}$} \\
 & \multicolumn{2}{c}{$\sqrt{s}=13.6$ TeV} & \multicolumn{2}{c}{$\sqrt{s}=14$ TeV} \\
Observable & $\sigma_{\mathcal{E}|\mathcal{B}}^{(0)}$ & $\sigma_{\mathcal{E}}^{\rm syst}$ & $\sigma_{\mathcal{E}|\mathcal{B}}^{(0)}$ & $\sigma_{\mathcal{E}}^{\rm syst}$ \\
\midrule
$V_{\rm ell}$ & 8.7 & 11 & 6.2 & 9.2 \\
$A_{\rm ell}$ & 19 & 52 & 4.9 & 41 \\
$\tan(2\theta_{rk})$ & 330 & 390 & 230 & 300 \\
$\mathcal{I}_{\rm QSE}^{(1)}$ & 5.3 & 19 & 1.2 & 15 \\
$\mathcal{I}_{\rm QSE}^{(2)}$ & 8.0 & 28 & 1.9 & 22 \\
$Q_{nn}$ & 2.9 & 11 & 1.0 & 8.9 \\
$Q_{rr}$ & 7.5 & 9.3 & 5.2 & 7.1 \\
$Q_{kk}$ & 7.9 & 19 & 5.2 & 15 \\
$Q_{rk}$ & 9.7 & 14 & 6.7 & 11 \\
$c_k$ & 11 & 16 & 8.7 & 13 \\
$c_r$ & 9.4 & 13 & 7.1 & 11 \\
$c_n$ & 110 & 100 & 75 & 74 \\
\end{tabular*}
\end{table}

\begin{figure}[t]
\centering
\includegraphics[width=\linewidth]{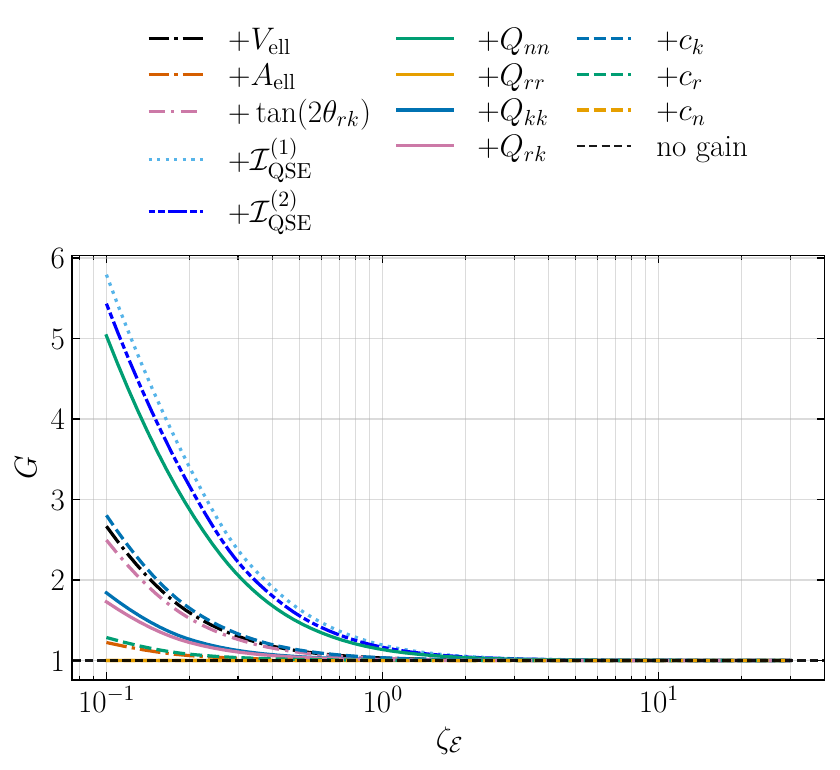}
\caption{Conditional-uncertainty scan for $\mathcal O_{\phi G}$ using the inclusive covariances of the $\lumi=300\,\rm fb^{-1}$ at $\sqrt{s}=13.6\,\rm TeV$ case. Each curve shows the gain obtained by adding one QSE observable $\mathcal E$ to the calibrated baseline in the $c_{\phi G}$ direction.}
\label{fig:cphiG_qse_gain_scan}
\end{figure}

The curves in Fig.~\ref{fig:cphiG_qse_gain_scan}, corresponding to the gain profile for $\mathcal O_{\phi G}$, show considerable gains beyond the nominal uncertainty, and only modest improvements at degraded precision. The most potent observables are both QSE Bell indicators. The other gluonic operators, $\OtG$ and $\mathcal O_G$, exhibit similar behavior, but also receive strong gains from $V_{\rm ell}$. For the four-fermion operators, both QSE Bell indicators remain the strongest observables, but $A_{\rm ell}$ stands out as particularly powerful, considerably more than the other observables.

\begin{figure}[!ht]
    \centering
    \includegraphics[width=\linewidth]{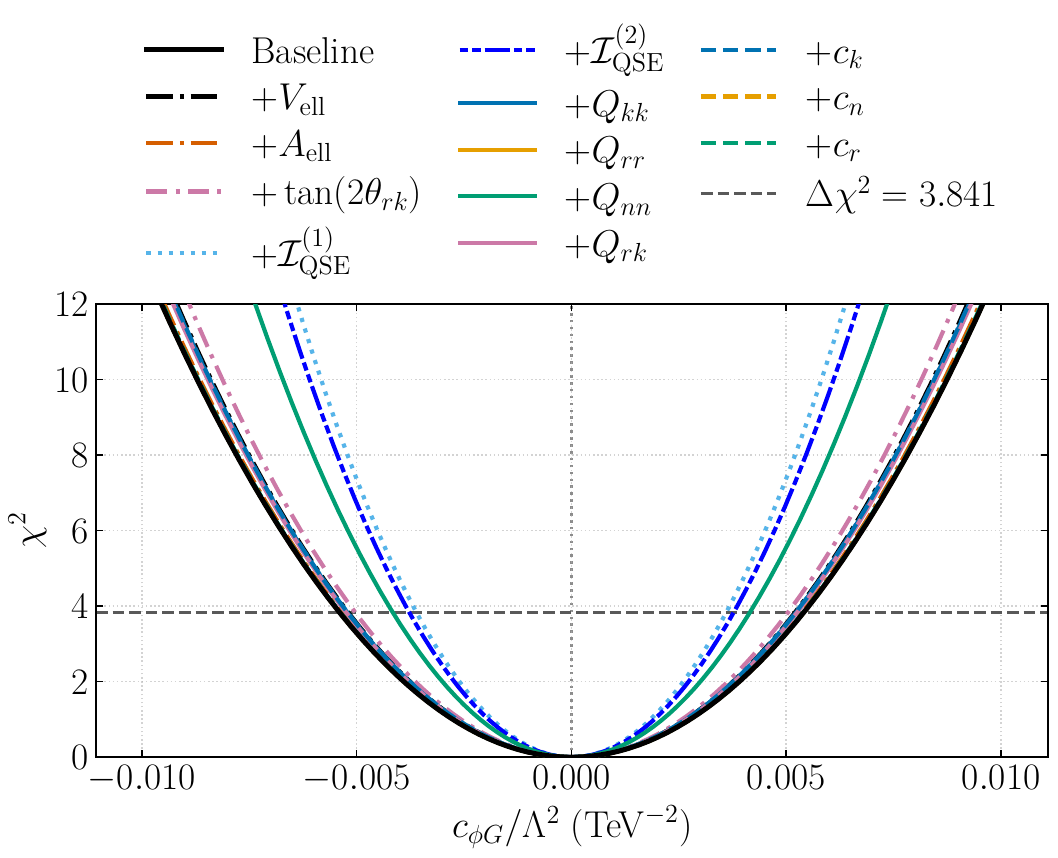}
    \caption{$\chi^2$ profile at $\zeta_\mathcal E=1$ for $\mathcal O_{\phi G}$ at $\lumi=300\,\rm fb^{-1}$ and $\sqrt{s}=13.6\,\rm TeV$. The crossing of the curves with the line at $\chi^2=3.841$ determines the constraint on the Wilson coefficient interval.}
    \label{fig:Profile_cphiG_plot}
\end{figure}

The reduction of the $c_{\phi G}$ interval is shown directly in Fig.~\ref{fig:Profile_cphiG_plot}. At nominal conditional precision, $\zeta_\mathcal E=1$, each curve represents the calibrated $\lumi=300\,\text{fb}^{-1}$ baseline supplemented by a single QSE observable. Since the same operator-dependent calibration is applied to all profiles, the displacement of the intersections with the 95\% confidence level isolates the additional conditional sensitivity carried by that observable. A steeper profile crosses the confidence threshold closer to the SM point and therefore yields a narrower 95\% confidence interval. Consistent with the gain scan in Fig.~\ref{fig:cphiG_qse_gain_scan}, the two QSE Bell indicators produce the largest contraction of the allowed region. In contrast, the remaining geometric observables produce more moderate improvements over the baseline. The profile representation thus makes explicit how the nonlinear QSE parameterization can reorganize the SMEFT response into directions that are more constraining after accounting for correlations with the measured Fano coefficients.

\subsubsection{Inclusive observable projections}
\label{subsubsec:inclusive_projections}

We now consider the selected QSE observables simultaneously and evaluate their combined impact using the inclusive $\lumi=300\,\rm fb^{-1}$ ($\sqrt{s} = 13.6$ TeV) and $\lumi=3000\,\rm fb^{-1}$ ($\sqrt{s} = 14$ TeV) samples. For each Wilson coefficient, we compare the calibrated CMS-motivated Fano baseline with the same baseline supplemented by the selected QSE observable set. The nominal results impose a reasonable correlation requirement of $|\rho_{ij}|\leq0.70$, while the observables selected with cutoffs of $0.80$ and $0.90$ are used to assess the stability of the projected gains. To avoid constructing combined likelihoods from nearly redundant observables, the QSE set is selected using the correlation matrix $\mathbf P_e$ associated with the covariance matrix $\mathbf V_e$, where $e\in\{\rm 300\,fb^{-1},3000\,fb^{-1}\}$ denotes the expected luminosities at different center-of-mass energies. Beginning with the fixed Fano baseline, a candidate observable $i$ is retained only when
\begin{align}
\max_{j\in\mathcal S}
\left|
(\mathbf P_e)_{ij}
\right|
&\leq
\rho_{\max},
&\frac{
\lambda_{\min}
\left(
\mathbf P_{e,\mathcal S\cup i}
\right)
}{
\lambda_{\max}
\left(
\mathbf P_{e,\mathcal S\cup i}
\right)
}\geq
10^{-2},
\end{align}
where $\mathcal S$ denotes the set already selected and
$\mathbf P_{e,\mathcal S\cup i}$ is the corresponding correlation submatrix. The nominal results use $\rho_{\max}=0.70$, while the values $0.80$ and $0.90$ define the stability envelope shown in the percentage-improvement figures.

\begin{figure*}[t]
\centering

\begin{minipage}[t]{0.49\textwidth}
    \centering
    \includegraphics[height=7.1cm]{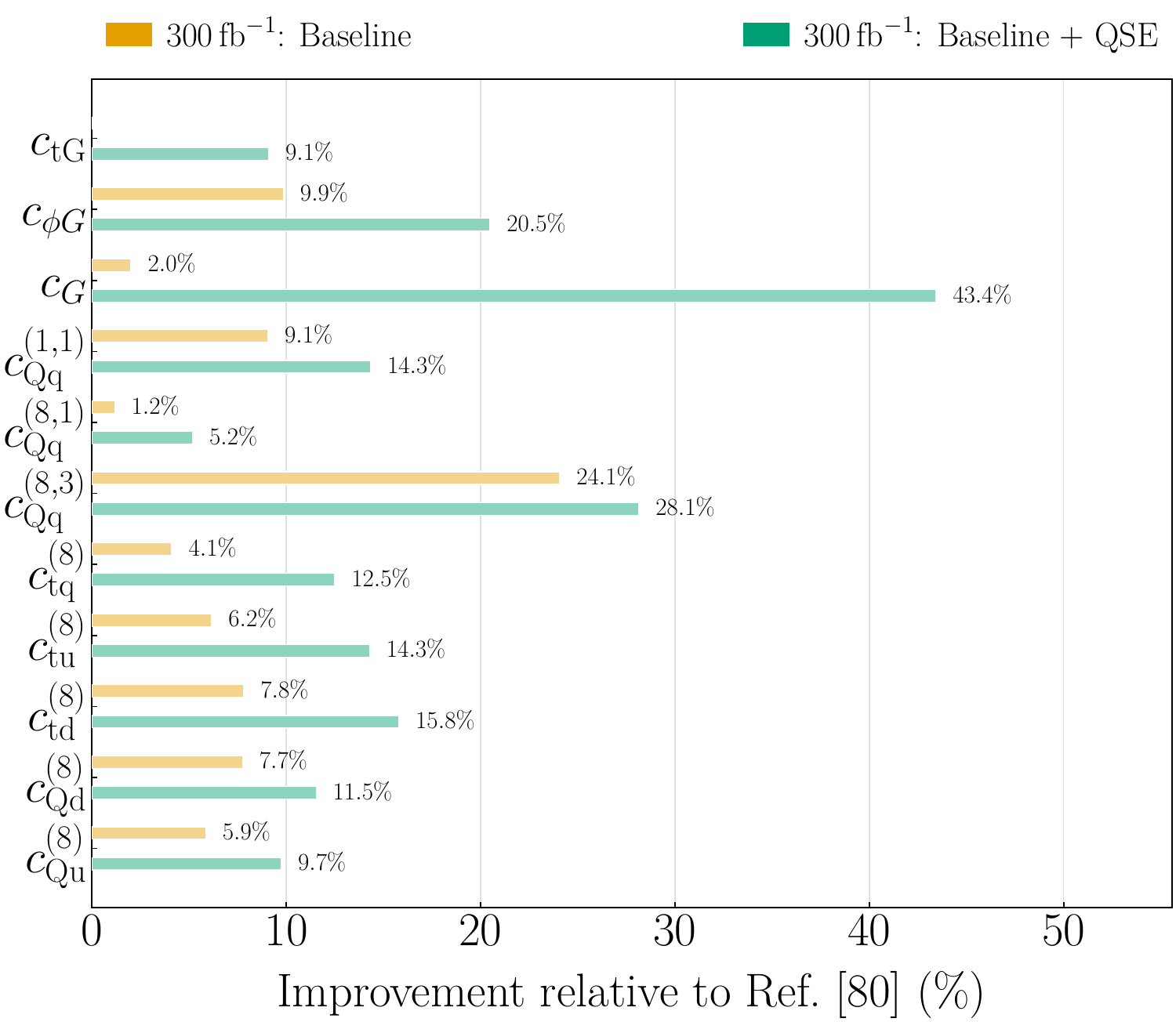}
\end{minipage}
\hfill
\begin{minipage}[t]{0.49\textwidth}
    \centering
\includegraphics[width=8cm,height=7.1cm]{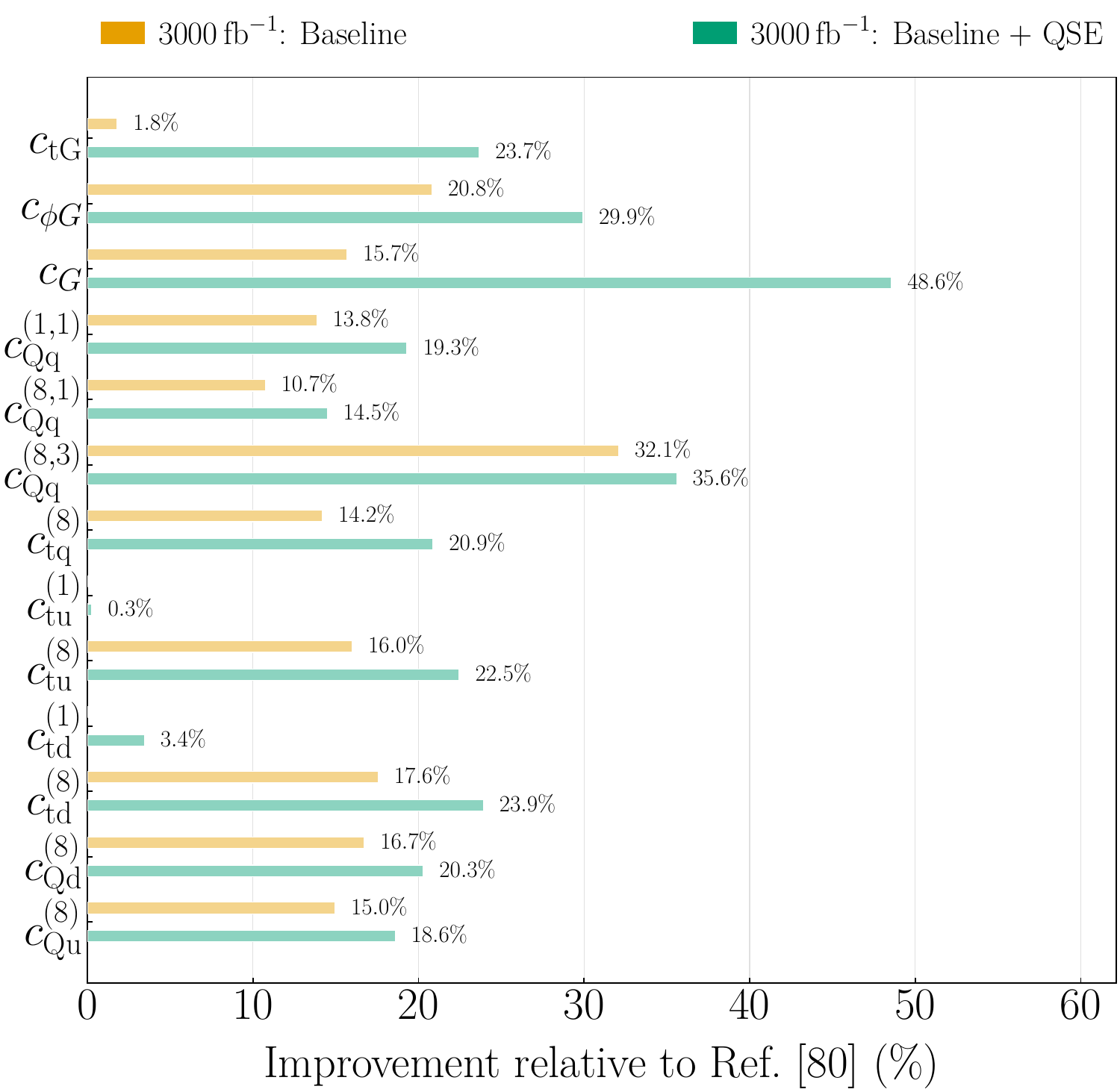}
\end{minipage}

\caption{One-coefficient $95\%$ confidence intervals for the inclusive projection at $\sqrt{s}=13.6\,\rm TeV$ with $\lumi=300\,\rm fb^{-1}$(left), and at $\sqrt{s}=14\,\rm TeV$ with $\lumi=3000\,\rm fb^{-1}$ (right). The published CMS intervals \cite{CMS:2019nrx} are compared with the calibrated CMS-motivated Fano baseline and the same baseline supplemented by the selected QSE observables. The projected intervals correspond to the nominal correlation selection $|\rho_{ij}|\leq0.70$, with all remaining Wilson coefficients set to zero.}
\label{fig:run3_hl_forest_limits}

\end{figure*}

The ``forest plots" in Fig.~\ref{fig:run3_hl_forest_limits} display the absolute projected sensitivity, while the incremental contribution of the QSE observables is quantified through the interval widths. For an observable set $\mathcal S$, we define $ w_j^{\mathcal S}\equiv c_{j,+}^{\mathcal S}-c_{j,-}^{\mathcal S}$. 
We quote the improvement from the QSE observables as the percentage reduction in the confidence-interval width relative to the corresponding Fano baseline.

\begin{figure}[!ht]
\centering
\includegraphics[width=\linewidth]{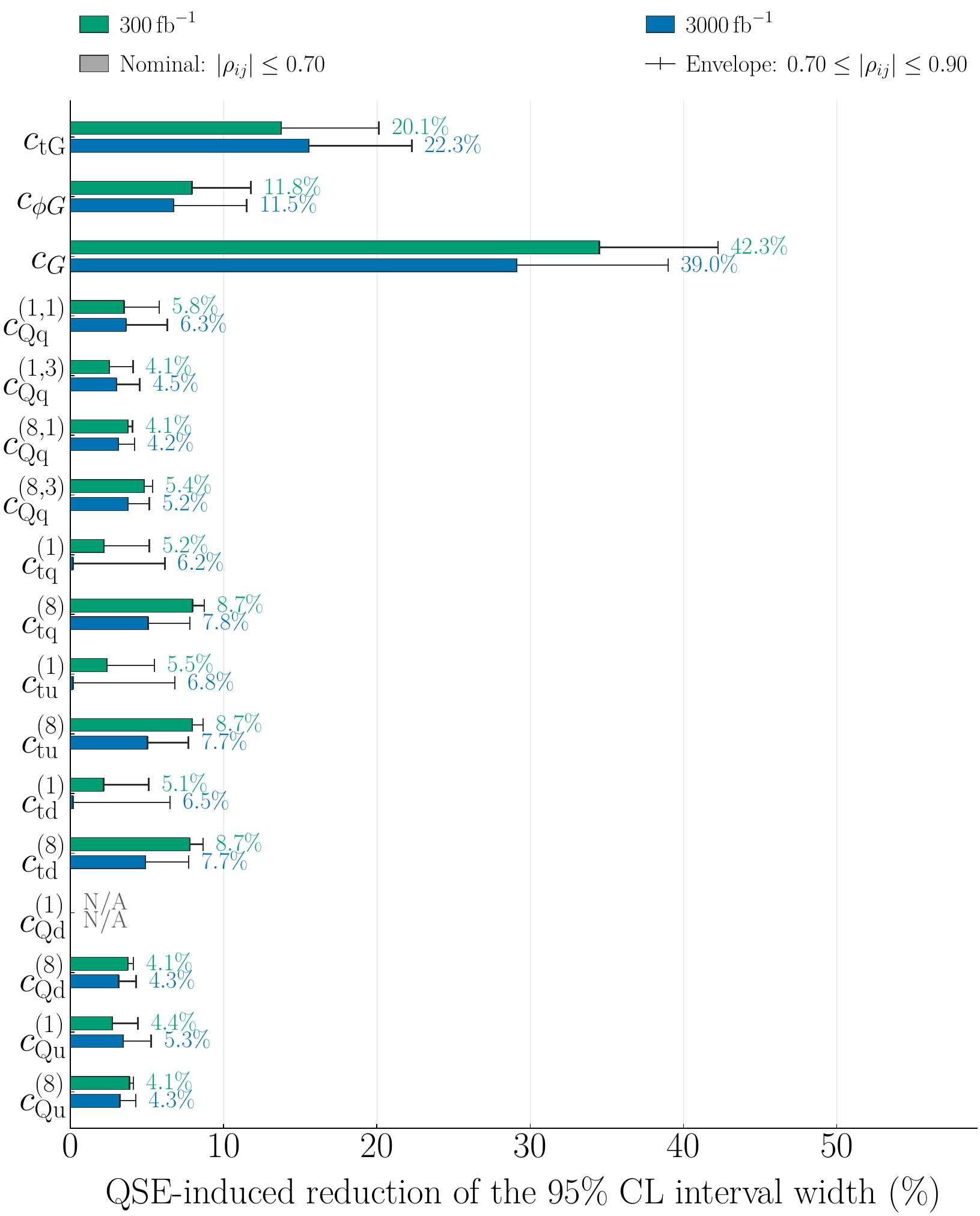}
\caption{QSE-induced reduction of the one-coefficient $95\%$ confidence-interval widths relative to the corresponding Fano baseline for $\sqrt{s}=13.6,\rm TeV$ with $\lumi=300\,\rm fb^{-1}$ (green), and $\sqrt{s}=14,\rm TeV$ with $\lumi=3000\,\rm fb^{-1}$ (blue),  
expressed as percentage reductions as explained earlier. The filled bars show the nominal selection $|\rho_{ij}|\leq0.70$, while the horizontal envelopes span the results obtained by varying the correlation cutoff between $0.70$ and $0.90$. The envelopes quantify the stability under the observable-selection prescription and do not represent additional experimental uncertainties.}
\label{fig:run3_hl_qse_gain}
\end{figure}

As shown in Fig.~\ref{fig:run3_hl_qse_gain}, the inclusion of the QSE observables narrows every 95\% interval determined for $\lumi = 300\,\rm fb^{-1}$. The most significant nominal reduction occurs for $c_G$, reaching $42.3\%$, followed by $20.1\%$ for $\ctG$ and $11.8\%$ for $c_{\phi G}$. The four-fermion directions receive smaller but systematic improvements, generally between $4.1\%$ and $8.7\%$, while $c_{\rm Qd}^{(1)}$ is not available.

At $\lumi=3000\,\rm fb^{-1}$, the relative QSE contribution remains strongly operator dependent. Again, the most sizeable reduction is obtained for $c_G$, at $39.0\%$, while the improvement for $\ctG$ increases to $22.3\%$ and that for $c_{\phi G}$ becomes $11.5\%$. The four-fermion gains are typically a few percent, reaching $7.7\%$ for $c_{\rm tu}^{(8)}$ and $7.7\%$ for $c_{\rm td}^{(8)}$, whereas some singlet directions receive only marginal nominal reductions. The QSE gain therefore does not scale uniformly with luminosity. As the Fano baseline becomes more precise, a smaller fractional improvement can still correspond to a substantial absolute contraction of the Wilson-coefficient interval.

To compare the projections directly with the observed CMS constraints for BSM, we also quote the percentage reduction in interval width relative to the observed CMS interval. This absolute improvement is distinguished from the incremental improvement obtained by adding QSE observables to the corresponding Fano baseline.
At $\lumi=300\,\rm fb^{-1}$, the ``baseline-plus-QSE" projection improves beyond the published CMS interval by as much as $43.4\%$ for $c_G$, $28.1\%$ for $c_{\rm Qq}^{(8,3)}$, and $20.5\%$ for $c_{\phi G}$, while keeping gains moderate for the rest of the coefficients. The ones that are not shown do not attain a positive gain relative to the experimental limits. At the $\lumi=3000\,\rm fb^{-1}$ level, the greater absolute sensitivity makes the projected improvements substantially broader: the combined intervals are narrower than the observed CMS limits by up to $48.6\%$ for $c_G$, $35.6\%$ for $c_{\rm Qq}^{(8,3)}$, and $29.9\%$ for $c_{\phi G}$. Large improvements are also obtained for $\ctG$, $c_{\rm tu}^{(8)}$, and $c_{\rm td}^{(8)}$, reaching $23.7\%$, $22.5\%$, and $23.9\%$, respectively.

These two comparisons therefore emphasize complementary aspects of the projection. The relative-to-baseline quantity isolates the additional information supplied by the QSE observables, whereas the comparison with CMS measures the absolute experimental reach. The results for $\lumi=3000\,\rm fb^{-1}$ show that the fractional QSE gain may decrease for some directions as the Fano baseline becomes more constraining. In contrast, the resulting combined intervals can nevertheless improve substantially over the currently published limits.

\section{Conclusion}

In this work, we present the first systematic application of quantum steering ellipsoids (QSEs) to collider physics, establishing their observables as a new class of ``geometric observables" for reconstructed quantum states and demonstrating their enhanced sensitivity to SMEFT effects.
QSEs are well established in quantum information science but have not previously been systematically applied to collider physics. We show that reconstructed spin tomography maps naturally onto the QSE formalism, enabling collider quantum states to be characterized through the geometry of their accessible conditional-state space.

Rather than introducing single new quantum information observables, this work establishes a ``geometric language" for collider quantum states. Conventional quantum information observables identify quantum-correlation thresholds, whereas QSE observables characterize the geometry of the same reconstructed quantum state. This complementary description reveals features of the reconstructed state that spin density and polarization observables do not directly capture, while remaining fully equivalent to the underlying spin density matrix.

Using $\ttbar$ production as a case study, we showed that different classes of SMEFT interactions produce distinct geometric signatures. Dipole operators primarily reshape the steering ellipsoid, whereas four-fermion operators additionally translate its center through induced single-spin polarizations. The QSE therefore separates different physical mechanisms into distinct geometric deformation channels, providing a direct geometric interpretation of how new physics modifies the reconstructed quantum state. Although mathematically equivalent to the reconstructed spin density matrix, the QSE reorganizes the same information into geometric observables with complementary and enhanced sensitivity to BSM interactions as implemented by SMEFT. Expected limits improve by anywhere from a few percent to 40\%. 

More broadly, this work establishes geometric quantum observables as a new framework for collider quantum information science. The present study demonstrates its phenomenological potential, while future work should quantify its performance in realistic complete experimental analyses, including detector effects, unfolding, the full set of systematic uncertainties, and differential measurements.

However, the framework is not limited to $\ttbar$ production. The approach developed here can be applied to any reconstructed bipartite system of spin-$1/2$ particles, and is therefore directly relevant to quantum state measurements at present and future $pp$, $ep$, $e^+e^-$, and $\mu^+\mu^-$ colliders. More generally, the underlying conditional-state construction suggests possible extensions to systems in which a qubit is coupled to a higher-dimensional subsystem, as well as to multipartite systems containing several qubits\cite{Jevtic:2014icx}. Developing the corresponding geometric observables and establishing their experimental reconstruction in such systems remain open questions. We anticipate that geometric descriptions of reconstructed quantum states can complement conventional quantum information observables in precision studies of the Standard Model and in searches for physics beyond it.

\section*{Acknowledgments}

We acknowledge support from the U.S. Department of Energy through the CMS High Energy ``Base Grant'' (Award No.~14000369) and the ``AI for a More Precise Future of the Top Quark'' project (Award No.~14000745).

OpenAI ChatGPT (GPT 5.5 and GPT 5.6 Sol) was used as an interactive tool to assist with the organization and revision of the manuscript and to refine the presentation of the theoretical formalism and scientific explanations. The authors directed its use and independently reviewed and verified all AI-assisted content. The authors assume full responsibility for the content of the manuscript.

\section*{Author Contributions}

J.J.M.A.: Formal analysis, Investigation, Methodology, Software, Validation, Visualization, Writing -- original draft, and Writing -- review \& editing.
A.J.W.: Conceptualization, Formal analysis, Methodology, Software, Supervision, Writing -- original draft, and Writing -- review \& editing.
A.A.: Investigation, Methodology, Software, Validation, Visualization, and Writing -- review \& editing.
J.M.D.-Q.: Investigation, Software, and Writing -- review \& editing.
S.B.: Validation and Writing -- review \& editing.
J.L.: Methodology, Supervision, and Writing -- review \& editing.
G.N.: Supervision and Writing -- review \& editing.
A.W.J.: Conceptualization, Funding acquisition, Methodology, Project administration, Resources, Supervision, and Writing -- review \& editing.

\appendix

\begin{widetext}

\section{Perturbative Spin Density Matrix}\label{app: Fano Coefficients DEF}

In Section~\ref{sec:smeft_qse_deformations} we introduced effective field theory deformations of the \ttbar quantum steering ellipsoid using perturbation theory to obtain the scattering amplitudes in terms of the Fano coefficients and, from there, compute the QSE orientation matrix. This appendix presents these analytical results up to order $\mathcal{O}(\Lambda^{-4})$. Everything is calculated at LO in QCD, with representative tree-level topologies being shown in Fig.~\ref{fig:analytical_pipeline_diagrams}.

\subsection{Operator Basis}

The relevant operators for our analysis are the zero- and two-fermion operators entering through the \gluglu channel:
\begin{equation}
\label{eq:two_fermi_ops}
\begin{aligned}
    \OG &= g_s f^{ABC} G^{A,\mu}_\nu G^{B,\nu}_\rho G^{C,\rho}_\mu\,, \\
    \OphiG &= \left( \phi^\dagger \phi - \frac{v^2}{2}\right) G_A^{\mu\nu}G^A_{\mu\nu}\,, \\
    \OtG &= g_s(\bar{Q}\sigma^{\mu\nu}T^A\,\Pqt)\tilde{\phi}G^A_{\mu\nu} + \text{h.c.}\,,
\end{aligned}
\end{equation}
where $\phi$ is the Higgs doublet and $\tilde\phi=\epsilon\phi$, $v$ is the Higgs vacuum expectation value (VEV), $g_s=\sqrt{4\pi\alpha_s}$ is the strong interaction coupling, $f^{ABC}$ and $T^A$ are the $\rm SU(3)$ structure constants and generators respectively, $\sigma^{\mu\nu}$ is the antisymmetric Dirac tensor, and $G^{\mu\nu}_A$ is the gluon field strength tensor. We also consider the color-octet and -singlet four-fermion operators acting on the \qqbar channel:
\begin{equation}
\label{eq:four_fermi_ops}
\begin{aligned}
    \cO_{Q\Pq}^{(8,1)} &= (\QbarL \gamma_\mu T^a \QL)(\overline{\text{q}}_L \gamma^\mu T^a \qL)\,, \\
    \cO_{Q\Pq}^{(8,3)} &= (\QbarL \gamma_\mu T^a \sigma^A \QL)(\overline{\text{q}}_L \gamma^\mu T^a \sigma^A \qL)\,,\\
    \cO_{\Pqt\Pqu}^{(8)} &= (\tbarR \gamma_\mu T^a \tR)(\ubarR \gamma^\mu T^a \uR) \,, \\
    \cO_{\Pqt\Pqd}^{(8)} &= (\tbarR \gamma_\mu T^a \tR)(\dbarR \gamma^\mu T^a \dR)   \,,\\
    \cO_{Q\Pqu}^{(8)} &= (\QbarL \gamma_\mu T^a \QL)(\ubarR \gamma^\mu T^a \uR)\,, \\
    \cO_{Q\Pqd}^{(8)} &= (\QbarL \gamma_\mu T^a \QL)(\dbarR \gamma^\mu T^a \dR)   \,,\\
    \cO_{\Pqt\Pq}^{(8)} &= (\tbarR \gamma_\mu T^a \tR)(\overline{\text{q}}_L \gamma^\mu T^a \qL) \,,
\end{aligned}
\end{equation}

Singlet operators can be obtained with the same expressions but without the $\rm SU(3)$ generators. The doublets of heavy and light quarks are $\QL$ and $\qL$, respectively, while the $\uR$ and $\dR$ are the right-handed light quarks.
We employ a typical flavor symmetry ${\rm U(2)_{\Pq} \otimes U(2)_{\Pqu} \otimes U(3)_{\Pqd}}$ to reduce the number of operators considered. 

\subsection{Fano Coefficients}

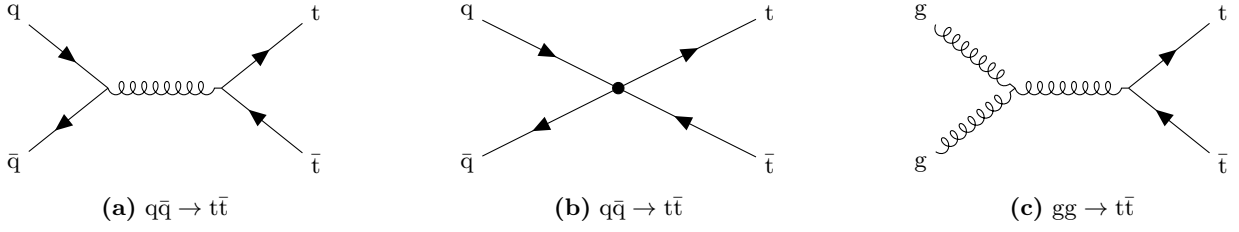
\begin{figure*}[t]
    \centering
    \begin{tikzpicture}
\begin{scope}[xshift=-6cm]
    \begin{feynman}
        \vertex (q)    at (-2, 1)   {$\rm q$};
        \vertex (qbar) at (-2,-1)   {$\rm \bar q$};

        \vertex (v1) at (-0.75,0);
        \vertex (v2) at ( 0.75,0);

        \vertex (t)    at (2, 1)    {$\rm t$};
        \vertex (tbar) at (2,-1)    {$\rm \bar t$};

        \diagram*{
            (q)    -- [fermion]      (v1),
            (qbar) -- [anti fermion] (v1),
            (v1)   -- [gluon] (v2),
            (v2)   -- [fermion]      (t),
            (v2)   -- [anti fermion] (tbar),
        };
    \end{feynman}

    \node at (0,-1.6)
    {\textbf{(a)} $\rm q\bar q\rightarrow t\bar t$};
\end{scope}

\begin{scope}[xshift=0cm]
    \begin{feynman}
        \vertex (q)    at (-2, 1)   {$\rm q$};
        \vertex (qbar) at (-2,-1)   {$\rm \bar q$};

        \vertex [dot] (v) at (0,0) {};

        \vertex (t)    at (2, 1)    {$\rm t$};
        \vertex (tbar) at (2,-1)    {$\rm \bar t$};

        \diagram*{
            (q)    -- [fermion]      (v),
            (qbar) -- [anti fermion] (v),
            (v)    -- [fermion]      (t),
            (v)    -- [anti fermion] (tbar),
        };
    \end{feynman}

    \node at (0,-1.6)
    {\textbf{(b)} $\rm q\bar q\rightarrow t\bar t$};
\end{scope}

\begin{scope}[xshift=6cm]
    \begin{feynman}
        \vertex (g1) at (-2, 1) {$\rm g$};
        \vertex (g2) at (-2,-1) {$\rm g$};

        \vertex (v1) at (-0.75,0);
        \vertex (v2) at ( 0.75,0);

        \vertex (t)    at (2, 1) {$\rm t$};
        \vertex (tbar) at (2,-1) {$\rm \bar t$};

        \diagram*{
            (g1) -- [gluon] (v1),
            (g2) -- [gluon] (v1),
            (v1) -- [gluon] (v2),
            (v2) -- [fermion]      (t),
            (v2) -- [anti fermion] (tbar),
        };
    \end{feynman}

    \node at (0,-1.6)
    {\textbf{(c)} $\rm gg\rightarrow t\bar t$};
\end{scope}

    \end{tikzpicture}

    \caption{
    Representative tree-level topologies entering the analytical
    calculation of the $\rm t\bar t$ spin density matrix.
    Panel~(a) shows the QCD quark--antiquark annihilation channel,
    panel~(b) shows a four-fermion SMEFT contact interaction, and
    panel~(c) shows the $s$-channel gluon-fusion topology. In addition, we also consider the $t$ and $u$ channels for the $\gluglu\to\ttbar$ process.
    Here $\rm q=u,d,s,c,b$ denotes an initial-state quark.
    }
    \label{fig:analytical_pipeline_diagrams}
\end{figure*}

We follow a typical expansion of the Fano coefficients we obtain from Eq.~\eqref{eq: R expansion in terms of Lambda} in the new physics scale $\Lambda$
\begin{equation}
\label{eq:SMEFT_expansion}
    X = X^{(0)} + \frac{1}{\Lambda^2} X^{(1)} + \frac{1}{\Lambda^4} X^{(2)} \,,
\end{equation}
where $X^{(0)}$ is the pure SM coefficient, $X^{(1)}$ is the linear contribution from interference between the dimension-six operators and the SM, and $X^{(2)}$ is the quadratic dimension-six contribution. All coefficients were calculated in \texttt{Mathematica}~\cite{Mathematica} using the \texttt{SmeftFR v3}~\cite{Dedes:2023zws} for Feynman rules generation, \texttt{FeynCalc 10}~\cite{Shtabovenko:2023idz} for the processing of scattering-related mathematics, and \texttt{FeynArts 3}~\cite{HAHN2001418} for visualization of Feynman diagrams. 
\subsection{Standard Model}

For the gluon-fusion channel, the non-vanishing LO QCD coefficients are
\begin{align*}
A^{\gluglu,(0)}
&=
F_{\gluglu}^{(0)}
\left[
1+2\beta^2(1-z^2)
-\beta^4(z^4-2z^2+2)
\right],
\\
\Cnn^{\gluglu,(0)}
&=
-F_{\gluglu}^{(0)}
\left[
1-2\beta^2+\beta^4(z^4-2z^2+2)
\right],
\\
\Ckk^{\gluglu,(0)}
&=
-F_{\gluglu}^{(0)}
\left[
1-2z^2(1-z^2)\beta^2
-(2-2z^2+z^4)\beta^4
\right],
\\
\Crr^{\gluglu,(0)}
&=
-F_{\gluglu}^{(0)}
\left[
1-(2-2z^2+z^4)\beta^2(2-\beta^2)
\right],
\\
\Crk^{\gluglu,(0)}
&=
2F_{\gluglu}^{(0)}\,
z(1-z^2)^{3/2}\beta^2\sqrt{1-\beta^2}.
\end{align*}
where the normalization factor is $F_{\gluglu}^{(0)}(\beta,z)=\frac{7+9\beta^2 z^2}{192(1-\beta^2 z^2)^2}$.\\
For the quark-antiquark channel, with $F_{\qqbar}^{(0)}(\beta,z)=\frac{1}{18}$, the non-vanishing LO QCD coefficients are

\begin{align*}
A^{\qqbar,(0)}
&=
F_{\qqbar}^{(0)}
\left[
2+\beta^2(z^2-1)
\right],
\\
\Cnn^{\qqbar,(0)}
&=
F_{\qqbar}^{(0)}
\beta^2(z^2-1),
\\
\Ckk^{\qqbar,(0)}
&=
F_{\qqbar}^{(0)}
\left[
\beta^2+z^2(2-\beta^2)
\right],
\\
\Crr^{\qqbar,(0)}
&=
F_{\qqbar}^{(0)}
\left[
2-\beta^2-z^2(2-\beta^2)
\right],
\\
\Crk^{\qqbar,(0)}
&=
2zF_{\qqbar}^{(0)}
\sqrt{(1-z^2)(1-\beta^2)}.
\end{align*}

\subsection{Linear interference terms}

The linear interference terms between the SM and SMEFT for the $\gluglu$ channel read:

\begin{align*}
A^{\gluglu,(1)}
&=
\frac{1}{\Lambda^2}
\frac{g_s^2}{1-\beta^2 z^2}
\Bigg[
\frac{g_s^2 v m_t(9\beta^2 z^2+7)}
     {12\sqrt{2}}\,\ctG
\\[-0.2em]
&\hspace{2.5cm}
-\frac{\beta^2 m_t^4}
       {4m_t^2-(1-\beta^2)m_h^2}\,c_{\phi G}
+\frac{9g_s^2\beta^2m_t^2z^2}{8}\,c_G
\Bigg],
\\[0.4em]
\Cnn^{\gluglu,(1)}
&=
\frac{1}{\Lambda^2}
\frac{g_s^2}{1-\beta^2 z^2}
\Bigg[
-\frac{7g_s^2 v m_t}{12\sqrt{2}}\,\ctG
\\[-0.2em]
&\hspace{2.5cm}
-\frac{\beta^2 m_t^4}
       {4m_t^2-(1-\beta^2)m_h^2}\,c_{\phi G}
+\frac{9g_s^2\beta^2m_t^2z^2}{8}\,c_G
\Bigg],
\\[0.4em]
\Crr^{\gluglu,(1)}
&=
\frac{1}{\Lambda^2}
\frac{g_s^2}{1-\beta^2 z^2}
\Bigg[
\frac{g_s^2 v m_t}
     {12\sqrt{2}\,(\beta^2 z^2-1)}
\Bigl[
-9\beta^4(z-z^3)^2
-7\beta^2(z^4-z^2+1)
+7
\Bigr]\ctG
\\[-0.2em]
&\hspace{2.5cm}
-\frac{\beta^2 m_t^4}
       {4m_t^2-(1-\beta^2)m_h^2}\,c_{\phi G}
+\frac{9g_s^2\beta^2m_t^2z^2}{8}\,c_G
\Bigg],
\\[0.4em]
\Ckk^{\gluglu,(1)}
&=
\frac{1}{\Lambda^2}
\frac{g_s^2}{1-\beta^2 z^2}
\Bigg[
\frac{g_s^2 v m_t(9\beta^2 z^2+7)}
     {12\sqrt{2}\,(\beta^2 z^2-1)}
\Bigl[
\beta^2(z^4-z^2-1)+1
\Bigr]\ctG
\\[-0.2em]
&\hspace{2.5cm}
+\frac{\beta^2 m_t^4}
       {4m_t^2-(1-\beta^2)m_h^2}\,c_{\phi G}
-\frac{9g_s^2\beta^2m_t^2z^2}{8}\,c_G
\Bigg],
\\[0.4em]
\Crk^{\gluglu,(1)}
&=
\frac{1}{\Lambda^2}
\frac{g_s^2}{1-\beta^2 z^2}
\Bigg[
\frac{g_s^2 v m_t\beta^2z(1-z^2)\Bigl[
9\beta^2
+(\beta^2-2)z^2
  \bigl(9\beta^2(z^2-1)+7\bigr)
-2
\Bigr]}
     {24\sqrt{2}\sqrt{(1-\beta^2)(1-z^2)}
      \,(\beta^2 z^2-1)}
\ctG
\\[-0.2em]
&\hspace{2.5cm}
+\frac{9g_s^2\beta^2m_t^2z}{8}
\sqrt{\frac{1-z^2}{1-\beta^2}}\,c_G
\Bigg].
\end{align*}

Meanwhile, for the $\qqbar$ channel we write down the Fano coefficients using a chirality-inspired basis, given by

\begin{align*}
    &c_{VV}^{(1),u}=\left(c_{\rm Qq}^{(1,1)}+c_{\rm Qq}^{(1,3)}+c_{\rm \Pqt u}^{(1)}+c_{\Pqt\rm q}^{(1)}+c_{\rm Qu}^{(1)}\right)/4 & c_{VA}^{(1),u}=\left(-c_{\rm Qq}^{(1,1)}-c_{\rm Qq}^{(1,3)}+c_{\Pqt u}^{(1)}-c_{\Pqt\rm q}^{(1)}+c_{\rm Qu}^{(1)}\right)/4\\
    &c_{AV}^{(1),u}=\left(-c_{\rm Qq}^{(1,1)}-c_{\rm Qq}^{(1,3)}+c_{\rm\Pqt u}^{(1)}+c_{\Pqt\rm q}^{(1)}-c_{\rm Qu}^{(1)}\right)/4 &c_{AA}^{(1),u}=\left(c_{\rm Qq}^{(1,1)}+c_{\rm Qq}^{(1,3)}+c_{\rm \Pqt u}^{(1)}-c_{\Pqt\rm q}^{(1)}-c_{\rm Qu}^{(1)}\right)/4
\end{align*}
where the corresponding combinations for the down-type quark are obtained by replacing $u\to d$ and the color-octet versions are obtained by replacing $(1)\to(8)$. With this convention we obtain,

\begin{align*}
A^{\qqbar,(1)}
&=
\frac{1}{\Lambda^2}
\frac{4g_s^2m_t^2}{9(1-\beta^2)}
\left[
\frac{\sqrt{2}g_s^2v}{m_t}(1-\beta^2)\ctG
+\left(2-(1-z^2)\beta^2\right)c_{VV}^{(8),u}
+2z\beta\,c_{AA}^{(8),u}
\right],
\\
\Cnn^{\qqbar,(1)}
&=
-\frac{1}{\Lambda^2}
\frac{4g_s^2m_t^2\beta^2(1-z^2)}{9(1-\beta^2)}
c_{VV}^{(8),u},
\\
\Crr^{\qqbar,(1)}
&=
\frac{1}{\Lambda^2}
\frac{4g_s^2m_t^2(1-z^2)}{9(1-\beta^2)}
\left[
\frac{\sqrt{2}g_s^2v}{m_t}(1-\beta^2)\ctG
+(2-\beta^2)c_{VV}^{(8),u}
\right],
\\
\Ckk^{\qqbar,(1)}
&=
\frac{1}{\Lambda^2}
\frac{4g_s^2m_t^2}{9(1-\beta^2)}
\left[
\left(\beta^2+z^2(2-\beta^2)\right)c_{VV}^{(8),u}
+2z\beta\,c_{AA}^{(8),u}
+\frac{\sqrt{2}g_s^2v}{m_t}z^2(1-\beta^2)\ctG
\right],
\\
\Crk^{\qqbar,(1)}
&=
-\frac{1}{\Lambda^2}
\frac{2g_s^2m_t^2}{9}
\sqrt{\frac{1-z^2}{1-\beta^2}}
\left[
\frac{\sqrt{2}g_s^2v}{m_t}(2-\beta^2)z\,\ctG
+4z\,c_{VV}^{(8),u}
+2\beta\,c_{AA}^{(8),u}
\right],
\\
B_{\rm r}^{\qqbar,(1)}
&=
-\frac{1}{\Lambda^2}
\frac{4g_s^2m_t^2}{9}
\sqrt{\frac{1-z^2}{1-\beta^2}}
\left[
\beta z\,c_{AV}^{(8),u}
+2c_{VA}^{(8),u}
\right],
\\
B_{\rm k}^{\qqbar,(1)}
&=
\frac{1}{\Lambda^2}
\frac{4g_s^2m_t^2}{9(1-\beta^2)}
\left[
\beta(1+z^2)c_{AV}^{(8),u}
+2z\,c_{VA}^{(8),u}
\right].
\end{align*}

Here we clearly observe that the only surviving four-fermion Wilson coefficients correspond to the color-octet operators. This is consistent with the color orthogonality between the SM induced scattering amplitude and the one induced by the color-singlet operators.

\subsection{Quadratic terms}

The dimension-six squared contributions of the $\gluglu$ channel read:

\begin{align*}
A^{\gluglu,(2)}&=\frac{1}{\Lambda^4}\frac{m_t^4}{1-\beta^2}\left[\frac{g_s^4v^2\left[9\beta^4z^4+4\beta^2(3z^2+4)-37\right]}{24m_t^2(\beta^2z^2-1)}\ctG^2+\frac{24\beta^2m_t^4}{\left[4m_t^2-(1-\beta^2)m_h^2\right]^2}c_{\varphi G}^2
+\frac{27g_s^4(1-\beta^2z^2)}{4(1-\beta^2)}c_G^2
\right.
\nonumber\\
&\hspace{2.4cm}
\left.
+\frac{
2\sqrt{2}\,\beta^2g_s^2vm_t(z^2-1)
}{
(\beta^2z^2-1)\left[4m_t^2-(1-\beta^2)m_h^2\right]
}\ctG c_{\varphi G}
+\frac{9g_s^4v}{2\sqrt{2}m_t}\ctG c_G
\right],
\\[0.em]
\Cnn^{\gluglu,(2)}
&=
\frac{1}{\Lambda^4}
\frac{m_t^4}{1-\beta^2}
\left[
\frac{
g_s^4v^2
\left[
9\beta^4z^2(z^2-2)
+2\beta^2(8z^2-13)
+19
\right]
}{
24m_t^2(\beta^2z^2-1)
}\ctG^2
+\frac{24\beta^2m_t^4}{\left[4m_t^2-(1-\beta^2)m_h^2\right]^2}c_{\phi G}^2
\right.
\nonumber\\
&\hspace{2.4cm}
\left.
+\frac{27g_s^4(\beta^2z^2-1)}{4(\beta^2-1)}c_G^2
+\frac{
2\sqrt{2}\,\beta^2g_s^2vm_t(z^2-1)
}{
(\beta^2z^2-1)\left[4m_t^2-(1-\beta^2)m_h^2\right]
}\ctG c_{\phi G}
+\frac{9g_s^4v}{2\sqrt{2}m_t}\ctG c_G
\right],
\\[0.5em]
\Crr^{\gluglu,(2)}
&=
\frac{1}{\Lambda^4}
\frac{m_t^4}{1-\beta^2}
\Bigg[
\\&\frac{
g_s^4v^2
}{
24m_t^2
}
\frac{
-9\beta^6z^2(z^4-2z^2+2)
+\beta^4(18z^6-57z^4+52z^2-14)
+\beta^2(28z^4-57z^2+58)
+18z^2-37
}{
(1-\beta^2z^2)^2
}\ctG^2
\nonumber\\
&\hspace{2.4cm}
\left.
+\frac{24\beta^2m_t^4}{\left[4m_t^2-(1-\beta^2)m_h^2\right]^2}c_{\phi G}^2
-\frac{27g_s^4\left[1-(2-\beta^2)z^2\right]}{4(1-\beta^2)}c_G^2
\right.
\nonumber\\
&\hspace{2.4cm}
\left.
+\frac{
2\sqrt{2}\,g_s^2vm_t\beta^2(1-z^2)
}{
(1-\beta^2z^2)\left[4m_t^2-(1-\beta^2)m_h^2\right]
}\ctG c_{\phi G}
-\frac{
9g_s^4v\left[1-(2-\beta^2)z^2\right]
}{
2\sqrt{2}m_t(1-\beta^2z^2)
}\ctG c_G
\right],
\\[0.5em]
\Ckk^{\gluglu,(2)}
&=
\frac{1}{\Lambda^4}
\frac{m_t^4}{1-\beta^2}
\Bigg[
\\&\frac{
g_s^4v^2
}{
24m_t^2
}
\frac{
9\beta^6z^2(z^4-2)
+\beta^4(-18z^6+25z^4-12z^2-14)
+\beta^2(-28z^4+81z^2+12)
-18z^2-19
}{
(1-\beta^2z^2)^2
}\ctG^2
\nonumber\\
&\hspace{2.4cm}
\left.
-\frac{24\beta^2m_t^4}{\left[4m_t^2-(1-\beta^2)m_h^2\right]^2}c_{\phi G}^2
+\frac{27g_s^4\left[1-(2-\beta^2)z^2\right]}{4(1-\beta^2)}c_G^2
\right.
\nonumber\\
&\hspace{2.4cm}
\left.
-\frac{
2\sqrt{2}\,g_s^2vm_t\beta^2(1-z^2)
}{
(1-\beta^2z^2)\left[4m_t^2-(1-\beta^2)m_h^2\right]
}\ctG c_{\phi G}
+\frac{
9g_s^4v\left[1-(2-\beta^2)z^2\right]
}{
2\sqrt{2}m_t(1-\beta^2z^2)
}\ctG c_G
\right],
\\[0.5em]
\Crk^{\gluglu,(2)}
&=
\frac{1}{\Lambda^4}
\frac{m_t^4}{\sqrt{1-\beta^2}}
\left[
\frac{
g_s^4v^2
}{
192m_t^2
}
\frac{
-144\beta^4z^3(1-z^2)^{3/2}
+16\beta^2z\sqrt{1-z^2}(14z^2-23)
+144z\sqrt{1-z^2}
}{
(1-\beta^2z^2)^2
}\ctG^2
\right.
\nonumber\\
&\hspace{2.4cm}
\left.
+\frac{27g_s^4z\sqrt{1-z^2}}{2(1-\beta^2)}c_G^2
-\frac{
2\sqrt{2}\,g_s^2vm_t\beta^2z\sqrt{1-z^2}
}{
(1-\beta^2z^2)\left[4m_t^2-(1-\beta^2)m_h^2\right]
}\ctG c_{\phi G}
+\frac{
9g_s^4vz\sqrt{1-z^2}
}{
\sqrt{2}m_t(1-\beta^2z^2)
}\ctG c_G
\right].
\end{align*}

Meanwhile, for $\qqbar$ we have:

\begingroup

\begin{align*}
A^{\qqbar,(2)}=&\frac{1}{\Lambda^4}\frac{m_t^4}{9(1-\beta^2)^2}\Biggl[\frac{g_s^4v^2}{m_t^2}(1-\beta^2)\bigl(2-\beta^2(1+z^2)\bigr)\ctG^2+\frac{4\sqrt{2}g_s^2v}{m_t}(1-\beta^2)\bigl(c_{VV}^{(8),u}+\beta z\,c_{AA}^{(8),u}\bigr)\ctG\\&+\beta^2\Biggl[9\bigl(c_{AA}^{(1),u}\bigr)^2(1+z^2)+2\bigl(c_{AA}^{(8),u}\bigr)^2(1+z^2)+z^2\Bigl(9\bigl(c_{AV}^{(1),u}\bigr)^2+2\bigl(c_{AV}^{(8),u}\bigr)^2+9\bigl(c_{VA}^{(1),u}\bigr)^2+2\bigl(c_{VA}^{(8),u}\bigr)^2+9\bigl(c_{VV}^{(1),u}\bigr)^2\\&+2\bigl(c_{VV}^{(8),u}\bigr)^2\Bigr)+9\bigl(c_{AV}^{(1),u}\bigr)^2+2\bigl(c_{AV}^{(8),u}\bigr)^2-9\bigl(c_{VA}^{(1),u}\bigr)^2-2\bigl(c_{VA}^{(8),u}\bigr)^2-9\bigl(c_{VV}^{(1),u}\bigr)^2-2\bigl(c_{VV}^{(8),u}\bigr)^2\Biggr]\\&+4\beta z\Bigl(9c_{AA}^{(1),u}c_{VV}^{(1),u}+2c_{AA}^{(8),u}c_{VV}^{(8),u}+9c_{AV}^{(1),u}c_{VA}^{(1),u}+2c_{AV}^{(8),u}c_{VA}^{(8),u}\Bigr)\\&+18\bigl(c_{VA}^{(1),u}\bigr)^2+4\bigl(c_{VA}^{(8),u}\bigr)^2+18\bigl(c_{VV}^{(1),u}\bigr)^2+4\bigl(c_{VV}^{(8),u}\bigr)^2\Biggr],\\
\displaybreak[3]
\Cnn^{\qqbar,(2)}=&\frac{1}{\Lambda^4}\frac{m_t^4\beta^2(1-z^2)}{9(1-\beta^2)^2}\Biggl[\frac{g_s^4v^2}{m_t^2}(1-\beta^2)\ctG^2+9\bigl(c_{AA}^{(1),u}\bigr)^2+2\bigl(c_{AA}^{(8),u}\bigr)^2+9\bigl(c_{AV}^{(1),u}\bigr)^2+2\bigl(c_{AV}^{(8),u}\bigr)^2-9\bigl(c_{VA}^{(1),u}\bigr)^2\\&-2\bigl(c_{VA}^{(8),u}\bigr)^2-9\bigl(c_{VV}^{(1),u}\bigr)^2-2\bigl(c_{VV}^{(8),u}\bigr)^2\Biggr],\\
\displaybreak[3]
\Crr^{\qqbar,(2)}=&\frac{1}{\Lambda^4}\frac{m_t^4(1-z^2)}{9(1-\beta^2)^2}\Biggl[\frac{g_s^4v^2}{m_t^2}(2-3\beta^2+\beta^4)\ctG^2+\frac{4\sqrt{2}g_s^2v}{m_t}(1-\beta^2)c_{VV}^{(8),u}\ctG\\&-\beta^2\Bigl(9\bigl(c_{AA}^{(1),u}\bigr)^2+2\bigl(c_{AA}^{(8),u}\bigr)^2+9\bigl(c_{AV}^{(1),u}\bigr)^2+2\bigl(c_{AV}^{(8),u}\bigr)^2+2\bigl(c_{VA}^{(8),u}\bigr)^2+9\bigl(c_{VV}^{(1),u}\bigr)^2+2\bigl(c_{VV}^{(8),u}\bigr)^2\Bigr)\\&-9(\beta^2-2)\bigl(c_{VA}^{(1),u}\bigr)^2+4\bigl(c_{VA}^{(8),u}\bigr)^2+18\bigl(c_{VV}^{(1),u}\bigr)^2+4\bigl(c_{VV}^{(8),u}\bigr)^2\Biggr],\\
\displaybreak[3]
\Ckk^{\qqbar,(2)}=&\frac{1}{\Lambda^4}\frac{m_t^4}{9(1-\beta^2)^2}\Biggl[\frac{g_s^4v^2}{m_t^2}(1-\beta^2)\bigl(z^2(2-\beta^2)-\beta^2\bigr)\ctG^2+\frac{4\sqrt{2}g_s^2v}{m_t}(1-\beta^2)z\bigl(\beta c_{AA}^{(8),u}+z c_{VV}^{(8),u}\bigr)\ctG\\&+\beta^2\Biggl[9\bigl(c_{AA}^{(1),u}\bigr)^2(1+z^2)+2\bigl(c_{AA}^{(8),u}\bigr)^2(1+z^2)+z^2\Bigl(9\bigl(c_{AV}^{(1),u}\bigr)^2+2\bigl(c_{AV}^{(8),u}\bigr)^2-9\bigl(c_{VA}^{(1),u}\bigr)^2-2\bigl(c_{VA}^{(8),u}\bigr)^2\\&-9\bigl(c_{VV}^{(1),u}\bigr)^2-2\bigl(c_{VV}^{(8),u}\bigr)^2\Bigr)+9\bigl(c_{AV}^{(1),u}\bigr)^2+2\bigl(c_{AV}^{(8),u}\bigr)^2+9\bigl(c_{VA}^{(1),u}\bigr)^2+2\bigl(c_{VA}^{(8),u}\bigr)^2+9\bigl(c_{VV}^{(1),u}\bigr)^2+2\bigl(c_{VV}^{(8),u}\bigr)^2\Biggr]+\\&\beta z\Bigl(9c_{AA}^{(1),u}c_{VV}^{(1),u}+2c_{AA}^{(8),u}c_{VV}^{(8),u}+9c_{AV}^{(1),u}c_{VA}^{(1),u}+2c_{AV}^{(8),u}c_{VA}^{(8),u}\Bigr)\\&+2z^2\Bigl(9\bigl(c_{VA}^{(1),u}\bigr)^2+2\bigl(c_{VA}^{(8),u}\bigr)^2+9\bigl(c_{VV}^{(1),u}\bigr)^2+2\bigl(c_{VV}^{(8),u}\bigr)^2\Bigr)\Biggr],\\
\displaybreak[3]
\Crk^{\qqbar,(2)}=&\frac{1}{\Lambda^4}\frac{2m_t^4}{9(1-\beta^2)}\sqrt{\frac{1-z^2}{1-\beta^2}}\Biggl[\frac{g_s^4v^2}{m_t^2}z(1-\beta^2)\ctG^2+\frac{\sqrt{2}g_s^2v}{m_t}\bigl(z(2-\beta^2)c_{VV}^{(8),u}+\\&\beta c_{AA}^{(8),u}\bigr)\ctG+\beta\Bigl(9c_{AA}^{(1),u}c_{VV}^{(1),u}+2c_{AA}^{(8),u}c_{VV}^{(8),u}+9c_{AV}^{(1),u}c_{VA}^{(1),u}+2c_{AV}^{(8),u}c_{VA}^{(8),u}\Bigr)\\&+z\Bigl(9\bigl(c_{VA}^{(1),u}\bigr)^2+2\bigl(c_{VA}^{(8),u}\bigr)^2+9\bigl(c_{VV}^{(1),u}\bigr)^2+2\bigl(c_{VV}^{(8),u}\bigr)^2\Bigr)\Biggr],\\
\displaybreak[3]
B_{r}^{\qqbar,(2)}=&\frac{1}{\Lambda^4}\frac{2m_t^4}{9}\sqrt{\frac{1-z^2}{(1-\beta^2)^3}}\Biggl[18c_{VA}^{(1),u}c_{VV}^{(1),u}+4c_{VA}^{(8),u}c_{VV}^{(8),u}\\&+z\beta\Bigl(9c_{AA}^{(1),u}c_{VA}^{(1),u}+2c_{AA}^{(8),u}c_{VA}^{(8),u}+9c_{AV}^{(1),u}c_{VV}^{(1),u}+2c_{AV}^{(8),u}c_{VV}^{(8),u}\Bigr)\Biggr]\\&+\frac{1}{\Lambda^4}2g_s^2vm_t^3\ctG\sqrt{\frac{2(1-z^2)}{(1-\beta^2)^3}}\Bigl[\beta z\,c_{AV}^{(8),u}+(2-\beta^2)c_{VA}^{(8),u}\Bigr],\\
\displaybreak[3]
B_{k}^{\qqbar,(2)}
={}&\frac{1}{\Lambda^4}\frac{2m_t^4}{9(1-\beta^2)^2}
\Biggl[
2z\Bigl(
9c_{VA}^{(1),u}c_{VV}^{(1),u}
+2c_{VA}^{(8),u}c_{VV}^{(8),u}
\Bigr)
+\beta(1+z^2)
\Bigl(
9c_{AA}^{(1),u}c_{VA}^{(1),u}
+2c_{AA}^{(8),u}c_{VA}^{(8),u}
\\
&\qquad
+9c_{AV}^{(1),u}c_{VV}^{(1),u}
+2c_{AV}^{(8),u}c_{VV}^{(8),u}
\Bigr)
+2z\beta^2
\Bigl(
9c_{AA}^{(1),u}c_{AV}^{(1),u}
+2c_{AA}^{(8),u}c_{AV}^{(8),u}
\Bigr)
\Biggr]
\\
&\quad
+\frac{1}{\Lambda^4}
\frac{2\sqrt{2}g_s^2vm_t^3\ctG}{9(1-\beta^2)}
\Bigl[
2z\,c_{VA}^{(8),u}
+\beta(1+z^2)c_{AV}^{(8),u}
\Bigr].
\end{align*}
\endgroup

We note that the deformation of the $\rm g\ttbar$ vertex induces interference between $\OtG$ and the four-fermion operators.

\end{widetext}

\bibliography{main}

\end{document}